\documentclass[%
 reprint,
 amsmath,amssymb,
 aps,
]{revtex4-2}

\usepackage{graphicx}
\usepackage{dcolumn}
\usepackage{bm}
\usepackage{physics}
\usepackage{placeins}
\usepackage{blindtext}
\usepackage{microtype}
\usepackage{ragged2e}
\usepackage[justification=RaggedRight]{caption}
\usepackage{subcaption}
\usepackage{microtype}

\usepackage{microtype}
\usepackage[english]{babel} 

\usepackage{xcolor}

\newtheorem{remark}{Remark}
\begin{document}

\preprint{APS/123-QED}

\title{An open harmonic chain: Exact vs global and local reduced dynamics}

\author{Melika Babakan$^{1,2}$, Fabio Benatti$^{2,3}$ and Laleh Memarzadeh$^{1}$}
\email{Corresponding author: memarzadeh@sharif.edu}
\affiliation{$^1$Physics Department, Sharif University of Technology, Tehran, Iran\\
$^2$Physics Department, University of Trieste, Trieste, Italy\\
$^3$Istituto Nazionale di Fisica Nucleare (INFN), Sezione di Trieste, Trieste, Italy}

\begin{abstract}
In the following, we
study the dissipative time-evolution of a quantum chain consisting of three coupled harmonic oscillators, the first and third of which weakly interact quadratically
with two independent thermal baths in equilibrium at different temperatures. 
Due to the quadratic form of the total Hamiltonian, the unitary dynamics of the compound system is formally analytically solvable and defines a one-parameter group of Gaussian maps which enables us to solve the exact dynamics of the chain numerically. 
Following the Gorini-Kossakowski-Sudarshan-Lindblad (GKSL) approach to open quantum systems,
one can perform the rotating wave approximation with respect to the interacting, or non-interacting chain Hamiltonian and respectively derive the so-called global and local master equations. The solutions of the ensuing different master equations can then be compared with the exact one possibly sorting out the two approaches in correspondence to different time-scales of the system. 
We derive the steady states of the open chain quantum dynamics in the two approaches and show that the behavior of fidelity between them versus inter-oscillator coupling depends on the two bath temperature, revealing the existence of a temperature-dependent critical inter-oscillator coupling strength that determines the domain of validity of each approach. When the newly found coupling is less than this critical value, the local approach outperforms the global approach, whereas for larger inter-oscillator coupling, the global approach is a better approximation of the exact evolution. This critical value of inter-oscillator coupling depends on the two bath temperatures which then play a crucial role  for deciding the best possible approximating open dynamics. 
\end{abstract}

\maketitle


\section{Introduction}
Unlike closed (isolated) quantum systems, open quantum systems interact with their environment, leading to richer, irreversible dynamics. Establishing a rigorous theoretical framework for describing the dynamics of open quantum systems is crucial for advancing our understanding of quantum mechanics, for dealing with quantum informational tasks and for fostering new quantum technologies.
Both from the theoretical and experimental sides, several and diverse settings require an attentive use of the paradigm of open quantum systems. 
Their list includes, among others, cold atoms  \cite{brantut_thermoelectric_2013}, ion traps \cite{barreiro_open-system_2011}, biological networks, light-harvesting complexes \cite{caruso_highly_2009, killoran_enhancing_2015}. Moreover, 
the behavior in time of fundamental features of quantum systems, such as quantum coherence \cite{Glauber-coherence-1963,Sudarshan-coherence-1963,Plenio-coherence-2014} and quantum entanglement \cite{Werner-entanglement-1989,Peres-entanglement-1996,HORODECKI-entanglement-1996} which are essential in quantum protocols, need to be described in the framework of open quantum systems. 
Also, the development of a quantum thermodynamic theory\cite{gemmer_quantum_2009,binder_thermodynamics_2018,deffner_quantum_2019,rice_irreversible_1978} or the investigations of quantum transport properties~\cite{davies_model_1978,R_Alicki_1979,Zoubi-2003,Rivas_2010,Migliore_2011,Werlang-Heat-transport-2015,Guimaraes-Nonequilibrium-quantum-chains-2016,Wichterich-heat-transport-2007,rivas_topological_2017,huelga_vibrations_2013} require as much precise description as possible of the actual dynamics of the quantum system under consideration. Hence, it is of vital importance to set up a firm basis for describing the dynamics of open quantum systems.

The master equation for a closed quantum system is given by the von-Neumann equation. 
When the system is open the evolution of the system and environment as a whole consists of a group of unitary maps which is generated by the total Hamiltonian. 
Due to the interaction between the system and the environment, the dynamics of the system itself is not unitary. By tracing out the environmental degrees of freedom, one obtains an integro-differential equation for the system's density matrix characterized by a very complex kernel. 
 In the weak-coupling limit, 
Gorini-Kossakowski-Sudarshan-Lindblad (GKSL) derived a master equation for system's density matrix, known as quantum Markov master equation \cite{gorini_completely_1976, lindblad_generators_1976}. They derived the form of generator of an irreversible, one parameter semi-group of linear maps in terms of the interaction Hamiltonian between the system and the environment. That generator guarantees the dynamic is given by a completely-positive trace-preserving (CPTP) map.
The original approach for derivation of the Markov master equation, called global approach, requires diagonalizing the 
system's Hamiltonian. For systems with interacting components, diagonalizing the 
system's Hamiltonian is not always straightforward or feasible. 
To overcome this challenge, the local approach is introduced. 
This approach is valid when the interaction between the system's components is weak compared to the system-environment coupling. In this case, one ignores the interactions within the  system, the remaining non-interacting Hamiltonian is diagonalized and its  eigen-projections are used for constructing the Markov master equation. 
Certainly, deriving  the master equation in the local approach is much easier compared to the global approach; however,  neglecting  the internal system interactions requires a thorough study of the validity of such a stance, about which a long debate has been developing~\cite{Zoubi-2003,Rivas_2010,Migliore_2011,levy_local_2014,Werlang-Heat-transport-2015,Guimaraes-Nonequilibrium-quantum-chains-2016,Landi-nonequilibrium-2016,Trushechkin_local-global-2016,Gerardo-local-global-2017,Hofer_local-global-2017,De-Chiara_2018,Cattaneo_2019,Giovannetti-local-global-2020,benatti_bath-assisted_2020,benatti_exact_2021,scali_local_2021,Lutz-local-global-2022}.
\begin{figure}
\centering
\includegraphics[width=0.8\linewidth]{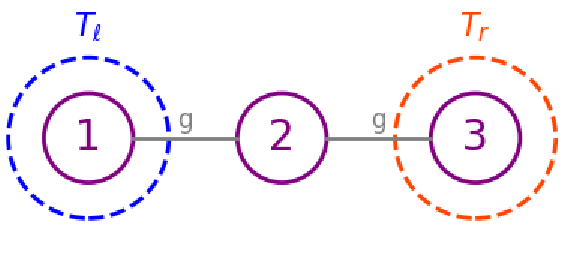}
\caption{The schematic representation of a three interacting harmonic oscillators which also interact with multi-mode external thermal baths at the two ends with temperatures $T_\ell$ and $T_r$.}
\label{fig:schematic representation}
\end{figure}

In full generality, to enhance one's grasp on the conditions of validity of either approach, one should compare their theoretical predictions with actual experimental evidences. However, conducting quantum experiments might result extremely difficult, especially in the presence of many internal degrees of freedom. Thus, we use an alternative method. 
We consider a solvable model for whole system and environment, which is a bosonic system-environment setting with a quadratic Hamiltonian. In this setting, the system is a chain of three interacting harmonic oscillators. 
At the two ends of this chain, the harmonic oscillators interact with two bosonic bath (See Fig.~(\ref{fig:schematic representation})). Each bath is a collection of non-interacting bosons which we assume to be in thermal states. The temperatures of the two baths are arbitrary.  
As the dynamics of the system-environment is exactly solvable, this setting enables us to exactly describe the density matrix of the system and environment at any arbitrary time. 
By tracing out over the environmental degrees of freedom, we obtain the system's density matrix at any time $t\geq 0$ in the so called exact approach. We compare this density matrix at time $t\geq 0$, with the solution of the Markov master equation derived in the Markov master equations appropriately derived in the global and in the local 
approaches, the comparison enabling us to assess their validity.

The proposed model, allows the straightforward derivation of the master equations in either approach. Indeed, even in the global approach it is not complicated to diagonalize the system's Hamiltonian because of its quadratic form. However, one has to take into account that the solution of the master equation in either approach is an operator on the infinite-dimensional Hilbert space.  
Such a difficulty is easily dealt with by noticing that a quadratic Hamiltonian generates a Gaussian dynamics which transforms Gaussian initial states into Gaussian states \cite{wolf-Gaussian-2006, OlivaresParis, eisert_gaussian_2005}(one calls such dynamics Gaussian). Because of the choice of the baths and of the system-bath interactions,
we will see that both the local and global open dynamics are Gaussian. Such an evidence makes it natural to concentrate on initial Gaussian states. Since they are completely identified by their so-called covariance matrices and displacement vectors, their dynamics will be equivalently assessed by the time-dependence of the latter two quantities.

We demonstrate the uniqueness of the steady state in either approach. In the global approach, we provide an analytical expression for the covariance matrix and the displacement vector of the steady state. In the local approach, we employ the perturbation theory to derive an analytical expression for the steady state to second order in the inter-oscillator coupling strength $g$. For arbitrary value of $g$, we compute the steady state of the local approach numerically. 
Our analysis reveals that in the limit of $g$ going to zero, the global steady state does not transform into the local steady state, unless the two baths have the same temperature and equal couplings to the chain. Comparison between the steady states of two master equations shows that when the inter-oscillator coupling is larger than a threshold value, by increasing $g$ the two steady states considerably depart form each other. The threshold value of $g$ depends on the two bath temperatures and vanishes when the latter are equal.

In order to determine which one between the  local and the global approach better describes the dynamics of the open quantum harmonic chain, we analyze the dynamics of their steady states 
under the exact evolution. The approach whose steady state depart less from itself under the exact evolution provides the more accurate description. This comparative study reveals a critical temperature-dependent inter-oscillator coupling strength $g_c$. For $g>g_c$, the global approach is preferable to the local one, while for $g<g_c$ the local approach outperforms the global one. This result establishes a neat criterion for selecting the appropriate approach based on the temperature of the baths, and on the inter-coupling strength in the chain.

The structure of the paper is as follows. After the necessary preliminaries in \S~\ref{sec:preliminaries}, in \S~\ref{sec:Model} we present the model.
In \S~\ref{sec:Exact approach}, we study the exact dynamics of the system initial Gaussian states under the full system-environment Hamiltonian. Section~\ref{sec:Disspative approach} is devoted to the derivation of the master equation in the local and global approaches. In \S~\ref{Sec:SteadyState}, we compute the local and global steady states which are then compared in~\S~\ref{sec:comparing-steady-state}, whereby 
the assessment of the superiority of one with respect to the other is also discussed. Finally, the results are summarized and commented emphasized
in~\S~\ref{sec:Conclusion}. 

\section{preliminaries}
\label{sec:preliminaries}
The first part of this section provides the necessary analytical tools to study the dynamics of Gaussian states under Gaussian semigroups. In the second part, we discuss the derivation of the Markov master equation in the local and global approach.

\subsection{Gaussian states and maps}
\label{sec:gaussian states}
In this subsection we establish our notation and review the definition of Gaussian states. Then we recall the properties of Gaussian dynamics and finalize the subsection with recalling Gaussian semigroups. 

Given a quantum system with $N$ bosonic modes with creation and annihilation operators $a^\dagger_i,a_i$, ($[a_i\,a^\dag_j]=\delta_{ij}\ ,\; [a_i\,,\,a_j]=[a^\dag_i\,,\,a^\dag_j]=0$), we introduce a vector $\hat{\xi}$ 
\begin{equation}
\label{eq:xidef}
    \hat{\xi}=(\xi_1,\xi_2,...,\xi_{2N})^{\top}
    =(a_1,a_2,...a_N,a_1^\dag,a_2^\dag,...,a_N^\dag)^{\top},
\end{equation}
with $\top$ denoting transposition. Its components satisfy the commutation relations
\begin{equation}
\label{eq:Omega}
    [\xi_i,\xi_j]=\Omega_{ij}  ,\;  \Omega=\begin{pmatrix}
        {\bf{0}}_N& \mathbb{I}_N\cr
        -\mathbb{I}_N & {\bf{0}}_N
    \end{pmatrix},
\end{equation}
where $\mathbb{I}_N$ denotes a $N$-dimensional identity matrix and ${\bf{0}}_N$ is a zero $N\times N$ matrix. 

In the following, we will focus upon a particular class of states called Gaussian states~\cite{serafini_quantum_2021}; they are completely 
identified by first, $\operatorname{Tr}(\rho \xi_i)=\langle\xi_i\rangle$, and second moments $\operatorname{Tr}(\rho \xi_i\xi_j)=\langle\xi_i\xi_j\rangle$. 

For any $N$-mode bosonic system with density operator $\rho$, the characteristic function is defined as follows:
\begin{equation}
 \label{eq:Zn}   \chi_\rho({Z})=\operatorname{Tr}(e^{-{{Z}}^\dagger \mathbb{Z}_N \hat{\xi}}\rho),\;\;\mathbb{Z}_N:=\begin{pmatrix}
        \mathbb{I}_N&\bf{0}_N\\
        \bf{0}_N&-\mathbb{I}_N
    \end{pmatrix},
\end{equation}
where $Z=(z_1,z_2,\cdots, z_N, z_1^*,z_2^*,\cdots z_N^*)^{\top}$ with $z_i\in\mathbb{C}$ being complex numbers. The corresponding various statistical moments are obtained by taking multiple derivatives of $\chi_\rho({Z})$ with respect to the components of $Z$ and then evaluating them at $Z=0$.
A state $\rho$ is said to be a Gaussian, if and only if its characteristic function is Gaussian:
\begin{equation}
    \chi_\rho({Z})=e^{-\frac{1}{4}{{Z}}^\dag C{{Z}} -i d^\top {Z}},
\end{equation}
where the covariance matrix $C$ and 
and the displacement vector $d$ are given in terms of the first and second moments of the state:
\begin{align}
    \label{eq:covaadag}
 C_{ij}=\langle\xi_i\xi_j^\dag+\xi_j^\dag\xi_i\rangle-2\langle\xi_i\rangle\langle \xi_j^\dag\rangle\ ,\quad d_i=\langle \xi_i\rangle\ .
\end{align}
While Gaussian states correspond to infinite-dimensional density operators, they can be completely described by a finite-dimensional covariance matrix and a displacement vector. 
By definition, the covariance matrix of $N$-mode Gaussian state is a positive matrix with the following general form
\begin{equation}
    C=\begin{pmatrix}
    \mathcal{C}_1& \mathcal{C}_2\\
    \mathcal{C}_2^\dagger & \mathcal{C}_1^{\top}
    \end{pmatrix},\;\mathcal{C}_1=\mathcal{C}_1^\dagger,
\end{equation}
where $\mathcal{C}_i$, $i=1,2$ are
$N\times N$matrices.
From Eq.~(\ref{eq:xidef}) and Eq.~(\ref{eq:covaadag}) it follows that the non-zero off-diagonal elements of $\mathcal{C}_{1,2}$ indicate the presence of
correlations between the corresponding degrees of freedom. For an explicit example, see Appendix~\ref{App:Exam}.

In general, because of interactions among their parties, initial Gaussian states of bosonic systems do not remain Gaussian during their time-evolution.
However, if the interactions are at most quadratic in annihilation and creation operators, these latter are dynamically mapped into linear combinations of themselves and Gaussian states remain Gaussian. Dynamics that preserve Gaussianity are called Gaussian dynamics. 
Such a preservation of Gaussianity occurs also in the case of  dissipative semigroup with generators consisting of Lindblad operators that are at most linear in annihilation and creation operators, called Gaussian semigroups~\cite{OlivaresParis}.
Therefore, Gaussian dissipative semigroups are identified by the dynamics of displacement vectors and covariance matrices.


\subsection{Markovian dynamics}
In this subsection, we outline the key steps for deriving Markov master equation in the so called weak-coupling limit. We recall the definition of the local and global approaches for derivation of Markov master equation. 

Master equations link the time derivative of quantum states to the action of a linear operator $\mathcal{L}$ known as the generator: $\dot{\rho}(t)=\mathcal{L}[\rho(t)]$.
In order to derive the generator for the dynamics of an open quantum system in interaction with its environment, the weak-coupling limit techniques (WCL) can be applied when 1) the coupling of system and environment is sufficiently weak, and 2) there is a clear separation between the time scale of the quantum system without environment  and the 
time-scale of the time-correlation functions of the latter, the weak coupling limit (WCL) allows to eliminate the environment and keep track of its presence in a generator that has a typical Gorini-Kossakowski-Sudarshan-Lindblad (GKSL) form. 
It can be derived from the exact evolution:
\begin{equation}
\label{eq:unitary}
\rho(0)\mapsto\operatorname{Tr}_B(U_{SB}(t)(\rho(0)\otimes\rho_B)U_{SB}^\dag(t)),
\end{equation}
by means of a number of approximations that ensures the generated maps
\begin{equation}
\Phi_t=\exp(t\mathcal{L}):\rho(0)\mapsto\rho(t),\;\;t\geq 0,
\end{equation}
are completely positive and trace preserving (CPTP). 
In Eq.~(\ref{eq:unitary}), $\operatorname{Tr}_B$ denotes the trace over the environment Hilbert space. The environment, typically modeled as a thermal bath, is described by the thermal state  $\rho_B$. The unitary dynamics, $U_{SB}(t)$ is generated by the total Hamiltonian of the compound system 
\begin{equation}
\hat{H}_{\rm tot}=\hat{H}_S+\hat{H}_B+\lambda \hat{H}_{\rm int}, 
\end{equation}
where $\lambda\ll1$ is a dimensionless 
coupling constant, $\hat{H}_S$ and $\hat{H}_B$ are respectively the system and 
bath Hamiltonians and $\hat{H}_{\rm int}$ is the interaction Hamiltonian between the system and the bath
\begin{equation}
\label{eq:intHamiltonian}
    \hat{H}_{\rm int}=\sum_iA_i\otimes B_i.
\end{equation}
The GKSL form of the master equation is as follows:
\begin{align}
    \dot{\rho}(t)&=\mathcal{L}[\rho(t)]=-i[\hat{H}_S+\lambda^2 \hat{H}_{\rm LS},\rho(t)]\cr
    &+\sum_{i,k}(A_i(\omega_k)\rho(t)A_i^\dag(\omega_k)-\frac{1}{2}\{A_i^\dag(\omega_k)A_i(\omega_k),\rho(t)\}),\cr
\end{align}
where $\hat{H}_{\rm LS}$ is the Lamb-shift Hamiltonian
\begin{equation}
    \hat{H}_{\rm LS}=\sum_{i,j,k}S_{ij}(\omega_k)A_j^\dag(\omega_k)A_i(\omega_k),
\end{equation}
with $S_{ij}(\omega_k)$ is 
\begin{equation}
    S_{ij}(\omega_k)=\mathcal{P}.\mathcal{V}\int_{-\omega_{max}}^{\omega_{\max}} d\nu \frac{\operatorname{Tr}\left(B_j^\dag(\nu)B_i\rho_B\right)}{\omega_k-\nu}.
\end{equation}
The Lindblad operators $A_i(\omega_k)$s are given by
\begin{equation}
\label{eq:LindOp}
    A_i(\omega_k)=\sum_{\epsilon-\epsilon'=\omega_k}P_{\epsilon'}A_iP_{\epsilon},
\end{equation}
where $P_{\epsilon}$ are projectors to the eigenstates of $\hat{H}_S$, and $A_i$ (acting on the system's Hilbert space) are defined in Eq.~(\ref{eq:intHamiltonian}). 

\begin{remark}
\label{rem1}
The global and local approaches to the derivation of the GKSL generator differ exactly in the choice of the eigenprojections of energy $P_{\epsilon}$.
While in the global approach they are chosen to be those of $\hat{H}_S$, in the local one they are instead those of the system Hamiltonian $\hat{H}^0_S$ where all interactions among the parties have been switched off. 
\end{remark}

\section{Model}\label{sec:Model}
In this section, we introduce our model and set our notation. Furthermore, we discuss under which circumstances the model can be solved analytically.

We consider a chain of three Harmonic oscillators with nearest neighbor interaction that also interacts with two thermal bath at the two ends of the chain (See Fig.~(\ref{fig:schematic representation})). Here, 
for simplicity, we restrict the discussion to a chain of three harmonic oscillators, although generalization to a chain of arbitrary size is straightforward. 
System Hamiltonian is given by
\begin{equation}
\label{eq:Hs}
	\hat{H}_S=\hat{\xi}_a^\dagger \mathsf{H}_S\hat\xi_a,
\end{equation}
where $\hat{\xi}_a$ is 
\begin{equation}
\label{eq:Xia}
    \hat{\xi}_a=(a_1,a_2,a_3,a_1^\dag,a_2^\dag,a_3^\dag)^\top.
\end{equation}
Here $a_i$ and $a_i^\dagger$ are respectively the annihilation and creation B‌osonic operators associated to harmonic oscillators of the same frequency $\omega_0>0$ at site $i$: $[a_i,a_j^\dagger]=\delta_{i,j}$, $[a_i,a_j]=[a_i^\dagger, a_j^\dagger]=0$. In Eq.~(\ref{eq:Hs}), $\mathsf{H}_S$
describes the free Hamiltonian of each oscillator and the interaction between nearest neighbors:
\begin{equation}
\label{eq:H_s}
    \mathsf{H}_{\rm S}=\frac{1}{2}(H_{\rm S}\oplus H_{\rm S}), \quad H_{\rm S}=\begin{pmatrix}
		\omega_0&g&0\cr
		g&\omega_0&g\cr
		0&g&\omega_0
	\end{pmatrix},
\end{equation}
 where $g>0$ is the inter-oscillator coupling. The Hamiltonian of the two baths at the ends of the chain is given by
\begin{equation}
	\hat{H}_B=\sum_{\alpha=\ell, r}\int_0^{+\infty} d\nu \;\nu b_{\alpha}^\dagger(\nu)b_{\alpha}(\nu),
\end{equation}
where $\alpha=\ell, r$ labels the left and right baths,  $b_{\alpha}(\nu)$
and $b_{\alpha}^\dagger(\nu)$ form a continuous family of Bosonic annihilation and creation operators for the continuous modes $\nu\geq 0$ in bath labeled by $\alpha$ : $[b_{\alpha}(\nu),b_{\beta}^\dagger(\nu')]=\delta_{\alpha,\beta}\delta(\nu-\nu')$. 
The two thermal baths and the harmonic oscillators at the two ends of the chain interact through a quadratic Hamiltonian of the form: 
\begin{align}
 \label{eq:Hint}
 \hat{H}_{\rm int}&=\int _{0}^{+\infty}d\nu h_\ell(\nu)\left( a_1^{\dagger}b_{\ell}(\nu)+a_1 b^{\dagger}_{\ell}(\nu)\right) \cr
	&+\int_{0}^{+\infty} d\nu h_r(\nu)\left( a_3^{\dagger}b_{r}(\nu)+a_3 b^{\dagger}_{r}(\nu)\right),
\end{align}
where $h_{\ell}(\nu)$ and $h_{r}(\nu)$ are the real coupling functions of the first and last harmonic oscillator to mode $\nu$ of the left and right baths.

In the following sections, we discuss the Gaussian dynamics of the system and environment, and also the one-parameter semigroup Gaussian evolution of the system, in more detail.  

\section{Exact approach}
\label{sec:Exact approach}
We start by discussing the exact unitary dynamics of the system and the two baths. In order to devise a dynamical setting amenable to numerical simulations, we consider a finite number of modes in each bath. Furthermore, we use the correspondence between unitary Gaussian dynamics and symplectic transformations. 

When the number of environment modes is finite and equal to $M$, the bath and interaction Hamiltonian become:
\begin{align}
\hat{H}_{\text{B}}&=\sum_{\alpha=\ell,r}\sum_{k=1}^M\omega_{\alpha, k}b_{\alpha,  k}^{\dagger}b_{\alpha, k},\cr
\hat{H}_{\text{int}}&=\sum_{k=1}^M h_{\ell, k}(a_1^{\dagger}b_{\ell, k}+a_1 b_{\ell, k}^{\dagger})\cr
&+\sum_{k=1}^M h_{r, k}(a_3^{\dagger}b_{r, k}+a_3 b_{r, k}^{\dagger}).
\end{align}
where $\omega_{\alpha,k}$ is the frequency of mode $k$ in the bath labelled by $\alpha$. The annihilation and creation operators of mode $k$ in bath $\alpha=\ell, r$ are respectively denoted by, $b_{\alpha,k}$
and $b_{\alpha,k}^\dagger$ for which we have 
$[b_{\alpha,k},b_{\beta,k}^\dagger]=\delta_{\alpha,\beta}\delta_{k,k'}$. The coupling constants of the 
first(last) Harmonic oscillator in the chain, to mode $k$-th of the left(right) become real numbers $h_{\ell,k}$  ($h_{r,k}$). Therefore, the total Hamiltonian of system and environment is summarised as follows:
\begin{equation}
\label{eq:HtotalDis}
\hat{H}_{\rm tot}=\hat{\xi}^\dagger \mathsf{H}_{\rm tot}\hat{\xi},
\end{equation}
with vectors of annihilation and creation operators enlarged to comprise also those of the two baths (see Eq.~(\ref{eq:xidef})): 
\begin{align}
&\hat{\xi}=(a_1,a_2,a_3,{\bf{b}}_\ell, {\bf{b}}_r, a^\dagger_1,a^\dagger_2,a^\dagger_3,{\bf{b}}^\dagger_\ell, {\bf{b}}^\dagger_r)^\top,\cr
&{\bf{b}}_{\alpha}=(b_{\alpha,1},b_{\alpha,2}\cdots,b_{\alpha,M})^{\top},\;\;\alpha=\ell ,r,
\end{align}
and
\begin{equation}
\mathsf{H}_{\rm tot}=\frac{1}{2}(\mathsf{H}\oplus\mathsf{H}),\quad \mathsf{H}=\begin{pmatrix}
		H_{\rm S}&\lambda L&\lambda R\cr
		\lambda L^\dagger&H_\ell&0\cr
		\lambda R^\dagger&0&H_{r}
	\end{pmatrix}.
\end{equation}
Here, $H_\alpha=\rm{diag}(\omega_{\alpha,1}, \omega_{\alpha,2},\cdots\omega_{\alpha,M})$ for $\alpha=\ell,r$ are diagonal matrices corresponding to the free Hamiltonian of left and right baths, $H_{\rm S}$ is defined in Eq.~(\ref{eq:H_s})
and the interaction between the system and environment are decoded in $L$ and $R$ which are $3\times M$ matrices:
\begin{equation}
	L=\begin{pmatrix}
		h_{\ell,1}&h_{\ell,2}&\cdots& h_{\ell.M}\cr
		0&0&\cdots& 0\cr
		0&0&\cdots& 0
	\end{pmatrix}, \;
	R=\begin{pmatrix}
		0&0&\cdots& 0\cr
		0&0&\cdots& 0\cr
		h_{r,1}&h_{r,2}&\cdots& h_{r,M}
		\end{pmatrix}.\\
\end{equation}
From the Bosonic commutation relations it follows that
\begin{equation}
[\hat{H}_{\rm tot},\hat{\xi}]=-2\mathbb{Z}_{2M+3}\mathsf{H}_{\rm tot}\hat{\xi},
\end{equation}
where $\mathbb{Z}_{2M+3}$ is defined in Eq.~(\ref{eq:Zn}).
Hence, in Heisenberg's picture,  the operators evolve as follows:
\begin{equation}
\hat{\xi}(t)=e^{it\hat{H}_{\rm tot}}\hat{\xi}e^{-it\hat{H}_{\rm tot}}=S\hat{\xi},
\end{equation}
the symplectic transformation $S$ is given by 
\begin{equation}
\label{eq:SymTran}
	S=e^{-2it\mathbb{Z}_{2M+3}\mathsf{H}_{\rm tot}}.
\end{equation}
Accordingly, the covariance matrix $C(t)$ and the displacement vector $d(t)$ at time $t$ are given by:
\begin{equation}
\label{eq:cdtransformation}
C_{\rm tot}(t)=SC_{\rm tot}(0)S^\dag, \;\;\; d(t)=Sd(0).
\end{equation}
When the harmonic chain  is initially prepared in the vacuum state and the left and the right baths are in equilibrium at temperatures $T_\ell$ and $T_r$, the initial covariant matrix and displacement vector are given by:
\begin{equation}
\label{eq:C0}
	C_{\rm tot}(0)=\begin{pmatrix}
		\mathcal{C}&{\bf{0}}\cr
		{\bf{0}}&\mathcal{C}
	\end{pmatrix},\;\; d(0)=0,
\end{equation}
with 
\begin{equation}
\mathcal{C}=\mathbb{I}_3\oplus\mathcal{C}_\ell\oplus\mathcal{C}_r,
\end{equation}
where $\mathcal{C}_\ell$ and $\mathcal{C}_r$ are the covariance matrices of the thermal states of the left and right bath:
\begin{equation}
({\bf{\mathcal{C}}}_{\alpha})_{i,j}=(2\bar{n}_{\alpha}(\omega_i)+1)\delta_{i,j},\;\; \forall i,j=[1,M],
\end{equation}
and the mean photon number of mode with frequency $\omega_k$ in bath $\alpha=\ell, r$ at temperature $T_\alpha$ in the natural units is given by 
\begin{equation}
\bar{n}_{\alpha}(\omega_k)=\frac{1}{e^{\beta_{\alpha}\omega_k}-1},\;\;\beta_{\alpha}=\frac{1}{T_{\alpha}}.
\end{equation}
We recall that, in the natural units Boltzman constant $k_B=1$. By using Eq.~(\ref{eq:SymTran}), Eq.~(\ref{eq:cdtransformation}) and Eq.~(\ref{eq:C0}) we have
\begin{equation}
	C_{\rm tot}(t)=\begin{pmatrix}
        e^{-it\mathsf{H}}\mathcal{C}e^{it\mathsf{H}}&{\bf{0}}\cr
        {\bf{0}}&e^{it\mathsf{H}}\mathcal{C}e^{-it\mathsf{H}}
	\end{pmatrix}, d(t)=0.
\end{equation}
This gives the exact covariance matrix of the system and the environment. The covariance matrix of the chain consists of the entries labeled with the indices of the three oscillators.

\section{Dissipative three harmonic oscillators}
\label{sec:Disspative approach}

In this section we discuss the dynamics of the chain of Harmonic oscillators in WCL and derive the master equations
in the two different approaches called ``\textit{local}" and ``\textit{global}" (see Remark~\ref{rem1}).
After deriving the master equation in either approach, we derive the differential equations for the covariance matrix and the displacement vector. 
         
Let us consider $N$-mode bosonic modes that evolve in time according to a master equation
\begin{equation}
\label{eq:general masterequation}
    \dot{\rho}=\mathcal{L}[\rho],
\end{equation}
with
\begin{equation}
\label{eq:generalgenerator}
    \mathcal{L}(\bullet)=-i[\hat{H},\bullet]+ \lambda^2\mathcal{D}[\bullet],
\end{equation}
with $\hat{H}$ being a quadratic Hamiltonian of $N$-mode bosonic system:
\begin{equation}
\label{eq:quadratic-Hamiltonian}
    \hat{H}=\hat{\xi}^\dagger \mathsf{H}\hat{\xi},
\end{equation}
and the dissipative part given in terms of Lindblad operators which are Bosonic creation and annihilation operators:
\begin{equation}
\label{eq:diss}
\mathcal{D}[\bullet]=\sum_{k=1}^{2N}\alpha_k(\xi_k^\dag\bullet\xi_k-\frac{1}{2}\{\xi_k\xi_k^\dag,\bullet\}),
\end{equation}
where $\alpha_k$ are real and positive.

To derive the equations of motion for the covariance matrix and the displacement vector, it is convenient to transform the master equation from the Schr\"odinger picture to the Heisenberg picture, $\dot{O}(t)=\mathcal{L}^{\rm adj}[O(t)]$ where $O(t)$ denotes the time-evolution up to time $t\geq 0$ of any  initial operator $O$.  The so-called adjoint of the generator in Eq.~(\ref{eq:generalgenerator}) is obtained by duality, namely through imposing the equality
\begin{equation}
\label{duality}
{\rm Tr}\Big(\rho_S\,\mathcal{L}^{\rm adj}[O]\Big)={\rm Tr}\Big(\mathcal{L}[\rho_S]\,O\Big),
\end{equation}
for all operators $O$ and states $\rho_S$ of the open Bosonic system $S$.
As shown in Appendix~\ref{sec:MasterCD}, using the expressions of the covariance matrix $C(t)$ and the displacement vector $d(t)$ in Eq.~(\ref{eq:covaadag}), we have:

\begin{align}
\label{eq:diffcovgen}
    \dot{C}(t)&=\mathcal{M}C(t)+C(t)\mathcal{M}^\dag+\mathcal{N},\\
    \label{eq:diffdisgen}
    \dot{d}(t)&=\mathcal{M}d(t),
\end{align}
where $\mathcal{M}$ and $\mathcal{N}$ are
\begin{align}
\label{eq:MN}
    \mathcal{M}&=i W+\frac{\lambda^2}{2}(D\mathbb{Z}_N+\mathbb{X}_ND\Omega),\\
   \mathcal{N}&=\lambda^2(D-\Omega D\Omega),\quad D_{ij}=\alpha_i \delta_{ij},\\
   W&=-\Omega(\mathsf{H}\mathbb{X}_N+\mathbb{X}_N\mathsf{H}),\quad \mathbb{X}_N=\begin{pmatrix}
        {\bf{0}}_N&\mathbb{I}_N\\
        \mathbb{I}_N&{\bf{0}}_N\\
    \end{pmatrix}.
\end{align}
$D$ is a $2N$ diagonal matrix with $\alpha_i$ in Eq.~(\ref{eq:diss}).
The solutions for the covariance matrix and displacement vector in Eq.~(\ref{eq:diffcovgen}) and Eq.~(\ref{eq:diffdisgen}) are \cite{Toscano-Thermal-equilibrium-Gaussian-dynamical-2022}:
\begin{align}
\label{solution1}
&C(t)=e^{t\mathcal{M}}C(0)e^{t\mathcal{M}^\dag}+ \int_0^t ds\,e^{s\mathcal{M}}\mathcal{N}e^{s\mathcal{M}^\dag},\\
&d(t)=e^{t\mathcal{M}}d(0),
\label{solution2}
\end{align}
where $C(0)$ and $d(0)$ are the initial covariance matrix and displacement vector at $t=0$, respectively.

\subsection{Local approach}
In this subsection, first, we derive the master equation in the local approach. Then we obtain the  
master equation for the system covariance matrix in the local approach. 

In the local approach, the Lindblad operators are given by Eq.~(\ref{eq:LindOp}) with the assumption that the inner-coupling between the Bosonic harmonic oscillator of the chain is zero. That means, the projectors in Eq.~(\ref{eq:LindOp}) are projectors to the eigenstates of $\hat{H}_S$ in Eq.~(\ref{eq:Hs}) when $g=0$. With this assumption, the master equation for system density matrix $\rho_S$ is given by
\begin{equation}
\dot{\rho}_S(t)=:\mathcal{L}_{\rm loc}[\rho_S(t)],
\end{equation}
with local generator
\begin{align}
\label{eq:masterlocal}
   \mathcal{L}_{\rm loc}[\rho_S(t)]&=-i[\hat{H}_S+\lambda^2\hat{H}^{\rm (LS)}_{\rm loc},\rho_S(t)]\cr
   &+2\pi\lambda^2\sum_{\alpha=l,r}\mathcal{D}_{\alpha,\rm loc}[\rho_S(t)],
\end{align}
with
\begin{align}
\mathcal{D}_{\ell,\rm loc}[\bullet]&=J_{\ell}(\omega_0)[\bar{n}_{\ell}(\omega_0)(a_1^{\dagger}\bullet a_1-\frac{1}{2}\{a_1a_1^{\dagger},\bullet\})\cr
    &+(\bar{n}_{\ell}(\omega_0)+1)(a_1\bullet a_1^{\dagger}-\frac{1}{2}\{a_1^{\dagger}a_1,\bullet\})],\\
\mathcal{D}_{r,\rm loc}[\bullet]&=J_{r}(\omega_0)\bar{n}_{r}(\omega_0)[(a_3^{\dagger}\bullet a_3-\frac{1}{2}\{a_3a_3^{\dagger},\bullet\})\cr
    &+(\bar{n}_{r}(\omega_0)+1)(a_3\bullet a_3^{\dagger}-\frac{1}{2}\{a_3^{\dagger}a_3,\bullet\})].
\end{align}
Here $J_{\alpha}(\omega_0)$ is the spectral density of either bath. Regardless of the constant term, the Lamb-Shift correction to the Hamiltonian in the local master equation Eq.~(\ref{eq:masterlocal}) is given by 
\begin{equation}
    \hat{H}^{\rm (LS)}_{ \rm loc}=\hat{\xi}_a^\dag \mathsf{H}^{\rm (LS)}_{\rm loc}  \hat{\xi}_a,
\end{equation}
where $\mathsf{H}^{\rm (LS)}_{\rm loc}$ is
\begin{align}
    \mathsf{H}^{\rm (LS)}_{\rm loc}&=\frac{1}{2}(H^{(\rm LS)}_{\rm loc}\oplus H^{(\rm LS)}_{\rm loc}),\\
    \label{eq:HLSloc}
    H^{(\rm LS)}_{\rm loc}&=\operatorname{diag}(S_{\ell}(\omega_0),0,S_{r}(\omega_0)).
\end{align}
 $S_{\alpha}(\omega)$ is a real-valued functions for all $\omega$ :
 \begin{equation}
 \label{eq:principalvalue}
    S_{\alpha}(\omega)=\mathcal{P}.\mathcal{V}\int_{0}^{\omega_{max}}d\omega' \frac{J_{\alpha}(\omega')}{\omega-\omega'},
\end{equation}
where $\omega_{max}$ is the cutoff frequency of the harmonic oscillators in the bath. In view of Eq.~(\ref{eq:diffcovgen}), the master equation in Eq.~(\ref{eq:masterlocal}) corresponds to the following equations of motions for 
covariance matrices and displacement vectors:
\begin{align}
    \label{eq:covlocal}
    &\dot{C}_{\rm loc}(t)=\mathcal{M}_{\rm loc}C_{\rm loc}(t)+C_{\rm loc}(t)\mathcal{M}_{\rm loc}^\dag+\mathcal{N}_{\rm loc},\\
    &\dot{d}_{\rm loc}(t)=\mathcal{M}_{\rm loc}d_{\rm loc}(0),
    \label{eq:displacementLocal}
\end{align}
with $\mathcal{M}_{\rm loc}$ and $\mathcal{N}_{\rm loc}$
\begin{align}
\label{eq:MNLocal}
    \mathcal{M}_{\rm loc}&=\mathsf{M}_{\rm loc}\oplus\mathsf{M}_{\rm loc}^\dagger,\\
    \mathcal{N}_{\rm loc}&=\mathsf{N}_{\rm loc}\oplus \mathsf{N}_{\rm loc},
    \label{eq:NLocal}
\end{align}
where 
\begin{align}
\label{eq:Mloc}
\mathsf{M}_{\rm loc}&=-i(H_{\rm S}+\lambda^2H^{\rm (LS)}_{\rm loc})-\pi\lambda^2\mathsf{J}_{\rm loc},\\
\label{eq:Nloc}
\mathsf{N}_{\rm loc}&=2\pi\lambda^2\mathsf{K}_{\rm loc}.\\
\mathsf{J}_{\rm loc}&=\operatorname{diag}(J_\ell(\omega_0),0,J_r(\omega_0)),\\
\mathsf{K}_{\rm loc}&=\operatorname{diag}(J_\ell(\omega_0)\tau_\ell(\omega_0),0,J_r(\omega_0)\tau_r(\omega_0)).
\end{align}
$H_{\rm S}$ and $H^{\rm (LS)}_{\rm loc}$ have been defined in Eq.~(\ref{eq:H_s}) and Eq.~(\ref{eq:HLSloc}) and
\begin{equation} 
\label{eq:taualpha}
\tau_\alpha(\omega_0)=2\bar{n}_\alpha(\omega_0)+1,\;\;\;\alpha=\ell,r.
\end{equation}
The solutions to Eq.~(\ref{eq:covlocal}) and Eq.~(\ref{eq:displacementLocal}) are obtained from Eq.~(\ref{solution1}) and Eq.~(\ref{solution2}) upon inserting $\mathcal{M}_{\rm loc}$ and $\mathcal{N}_{\rm loc}$ in the place of $\mathcal{M}$ and $\mathcal{N}$.

\subsection{Global approach}
In this subsection, we derive the master equation in the global approach after diagonalizing the system Hamiltonian. Then we derive the master equation for the system's covariance matrix. 

As already discussed, in the global approach the diagonalization of the system Hamiltonian is essential and it can be achieved as follows.
We introduce the vector of annihilation and creation operators: 
\begin{equation}
\label{eq:xiglb}
    \hat{\xi}_c=(c_1,c_2,c_3,c_1^\dag,c_2^\dag,c_3^\dag)^\top\ ,
\end{equation}
that is obtained by transforming $\hat{\xi}_a$ via the linear transformation
\begin{equation}
\label{eq:basis}
    \hat{\xi}_c=T\hat{\xi}_a\ ,\ T=\mathsf{T}\oplus \mathsf{T}\ ,\quad \mathsf{T}=\frac{1}{2}\begin{pmatrix}
        1&-\sqrt{2}&1\\
        \sqrt{2}&0&-\sqrt{2}\\
        1&\sqrt{2}&1\\
    \end{pmatrix}\ .
\end{equation}
Then, the Hamiltonian in Eq.~(\ref{eq:Hs}) reads
\begin{equation}
\label{eq:Hsglb}
	\hat{H}_S=\hat{\xi}_c^\dagger \mathsf{H}_d\hat\xi_c\ ,
\end{equation}
where
\begin{equation}
\label{eq:H_sglb}
\mathsf{H}_{\rm d}=\frac{1}{2}(H_{\rm d}\oplus H_{\rm d}),\quad H_{\rm d}=\begin{pmatrix}
		\epsilon_1&0&0\cr
		0&\epsilon_2&0\cr
		0&0&\epsilon_3
	\end{pmatrix},
\end{equation}
is a diagonal matrix with eigenvalues
\begin{equation}
    \epsilon_1=\omega_0-\sqrt{2}g,\;\;\;
    \epsilon_2=\omega_0,\;\;
    \epsilon_3=\omega_0+\sqrt{2}g\ .
\end{equation}
Therefore, the chain Hamiltonian becomes diagonal in the new annihilation and creation operators with eigen-energies 
$\sum_{i=1}^3 n_i\epsilon_i$, $n_i\in\mathbb{N}$ and  eigenvectors 
\begin{equation}
    \ket{n_1,n_2,n_3}=\frac{(c_1^\dag)^{n_1}(c_2^\dag)^{n_2}(c_3^\dag)^{n_3}}{\sqrt{(n_1)!(n_2)!(n_3)!}}\ket{vac},
\end{equation}
where $\ket{vac}$ denotes the vacuum vector.

Among all possible transitions, the system-bath interaction in Eq.~(\ref{eq:Hint}) only allows transition frequencies equal to $\pm\epsilon_i$, for $i=1,2,3$ when each $\epsilon_i$ is non-negative. Given that $\omega_0>0$ and $g>0$, demanding $\epsilon_i>0,\;\forall i$, establishes an upper bound on the coupling strength $g$ for physical solutions: $0<g<\frac{\omega_0}{\sqrt{2}}$. 

The ultimate master equation for the system density operator in the global approach is derived to be 
\begin{equation}
\label{eq:masterglobal}
\dot{\rho}_S=:\mathcal{L}_{\rm glb}[\rho_S(t)],
\end{equation}
with
\begin{align}
\label{eq:LGlobal}
    \mathcal{L}_{\rm glb}[\rho_S]&=-i[\hat{H}_S+\lambda^2\hat{H}^{\rm (LS)}_{\rm glb},\rho_S(t)]\cr
    &+2\pi\lambda^2\sum_{i=1}^3\mathcal{D}^{(i)}_{\rm glb}[\rho_S(t)].
\end{align}
Here, the Lamb-Shift correction to the Hamiltonian is given by:
\begin{equation}
    \hat{H}^{\rm (LS)}_{ \rm glb}=\hat{\xi}_c^\dag \mathsf{H}^{\rm (LS)}_{\rm glb}  \hat{\xi}_c,
\end{equation}
with 
\begin{align}
    \mathsf{H}^{\rm (LS)}_{\rm glb}&=\frac{1}{2}(H^{(\rm LS)}_{\rm glb}\oplus H^{(\rm LS)}_{\rm glb}),\\
    \label{eq:HLSglb1}
    H^{(\rm LS)}_{\rm glb}&=\operatorname{diag}\big(S(\epsilon_1),S(\epsilon_2),S(\epsilon_3)\big).
\end{align}
The real-valued function $S(\epsilon_i)$ reads as
\begin{equation}
    \label{eq:Sglobal}S(\epsilon_i)=\sum_{\alpha=\ell,r}S_{\alpha}(\epsilon_i),
\end{equation}
where $S_{\alpha}(\epsilon_i)$ is defined in Eq.~(\ref{eq:principalvalue}).
The dissipative terms in Eq.~(\ref{eq:LGlobal}) are 
\begin{align}
    \mathcal{D}^{(i)}_{\rm glb}[\bullet&]=\frac{1}{4}\sum_{\alpha=\ell,r}J_{\alpha}(\epsilon_i)[\bar{n}_{\alpha}(\epsilon_i)(c_i^{\dagger}\bullet c_i-\frac{1}{2}\{c_ic_i^{\dagger},\bullet\})\nonumber\\
    &+(\bar{n}_{\alpha}(\epsilon_i)+1)(c_i\bullet c_i^{\dagger}-\frac{1}{2}\{c_i^{\dagger}c_i,\bullet\})],\;\;i=1,2,\nonumber\\
    \mathcal{D}^{(2)}_{\rm glb}[\bullet]&=
\frac{1}{2}\sum_{\alpha=\ell,r}J_{\alpha}(\epsilon_2)\Big[\bar{n}_{\alpha}(\epsilon_2)(c_2^{\dagger}\bullet c_2-\frac{1}{2}\{c_2c_2^{\dagger},\bullet\})
\nonumber\\  
&+(\bar{n}_{\alpha}(\epsilon_2)+1)(c_2\bullet c_2^{\dagger}-\frac{1}{2}\{c_2^{\dagger}c_2,\bullet\})\Big].
\label{globalLind2}
\end{align}

In the global approach, the dynamics of the covariance matrix and of the displacement vector are obtained from Eq.~(\ref{solution1}) 
and Eq.~(\ref{solution2}) by inserting in place of $\mathcal{M}$ and $\mathcal{N}$ the matrices 

\begin{align}
\label{eq:MNGlobal}
    \widetilde{\mathcal{M}}_{\rm glb}&=\widetilde{\mathsf{M}}_{\rm glb}\oplus\widetilde{\mathsf{M}}_{\rm glb}^\dagger,\\
    \widetilde{\mathcal{N}}_{\rm glb}&=\widetilde{\mathsf{N}}_{\rm glb}\oplus \widetilde{\mathsf{N}}_{\rm glb},
    \label{eq:NGlobal}
\end{align}
where 
\begin{align}
\label{eq:Mglobal1}
&\widetilde{\mathsf{M}}_{\rm glb}=-i(H_{\rm d}+\lambda^2H^{\rm (LS)}_{\rm glb})-\frac{\pi}{4}\lambda^2\tilde{\mathsf{J}}_{\rm glb},\\
\label{eq:Nglobal1}
&\widetilde{\mathsf{N}}_{\rm glb}=\frac{\pi}{2}\lambda^2\Tilde{{\mathsf{K}}}_{\rm glb},\\
&\widetilde{\mathsf{J}}_{\rm glb}=\sum_{\alpha=l,r}\tilde{\mathsf{J}}_{\rm glb, \alpha},\\
&\tilde{\mathsf{K}}_{\rm glb}=\sum_{\alpha=\ell,r}\tilde{\mathsf{J}}_{\rm glb,\alpha}\tilde{\mathsf{K}}_{\rm glb \alpha}.
\end{align}
$H_{\rm d}$ and $H^{\rm (LS)}_{\rm glb}$ have been defined in Eq.~(\ref{eq:H_sglb}) and Eq.~(\ref{eq:HLSglb1}). Furthermore, $\widetilde{\mathsf{J}}_{\rm glb,\alpha}$ and $\widetilde{\mathsf{K}}_{\rm glb,\alpha}$ are defined as follows
\begin{align}
    \widetilde{\mathsf{J}}_{\rm glb,\alpha}&=\operatorname{diag}(J_{\alpha}(\epsilon_1),2J_{\alpha}(\epsilon_2)J_{\alpha}(\epsilon_3)),\\
    \Tilde{{{\mathsf{K}}}}_{\rm glb,\alpha}&=\operatorname{diag}(2\bar{n}_{\alpha}(\epsilon_1)+1,2\bar{n}_{\alpha}(\epsilon_2)+1,2\bar{n}_{\alpha}(\epsilon_3)+1).
\end{align}

\begin{remark}
\label{rem2}
The covariance matrix $\widetilde{C}_{\rm glb}(t)$ and the displacement vector $\widetilde{d}_{\rm glb}(t)$ refer to the annihilation and creation operators $c_i,c^\dag_i$; to each one of these pairs there correspond commuting contributions to the generator of the dissipative dynamics in the global approach. Then,
$\mathcal{L}_{\rm glb}$ splits into the sum of three commuting generators, $\mathcal{L}_{\rm glb}= \sum_{j=1}^3\mathcal{L}_j$. Consequently, the generated 
dynamics factorizes, 
\begin{equation}
\label{3factors}
\exp\Big(t\,\mathcal{L}_{\rm glb}\Big)=\prod_{j=1}^3\exp\Big(t\,\mathcal{L}_j\Big).
\end{equation}
In order to express the dynamics in terms of the annihilation and creation operators $a_i,a^\dag_i$, one uses the transformation matrix $T$ in Eq.~(\ref{eq:basis}) and recast the covariance matrix and displacement vector as 
\begin{align}
\label{eq:covtrasformedmatrix}
    C_{\rm glb}(t)=T^\dag\widetilde{C}_{\rm glb}(t)T,\quad 
    d_{\rm glb}(t)=T^\dag\widetilde{d}_{\rm glb}(t)\ .
\end{align}
\end{remark}
In the next section, we discuss the steady state of either approach. 
\section{steady state}
\label{Sec:SteadyState}
In this section, we discuss the steady state of the master equations given by local and global approaches. First, we discuss the existence and the uniqueness of the steady state. Then we derive the form of the steady states in either approach. 

The existence of the steady state follows since the time-average of any initial state evolving under either semigroup exists and is time-invariant.
Furthermore, since the only chain operators 
commuting with all Lindblad operators and with the Hamiltonian are scalar multiples of the identity, there can be only one invariant state to which all initial states converge in time~\cite{spohn_algebraic_1977}.
This is the case for both the local and global generators discussed above, for only the identity operator commutes with the system Hamiltonian 
in Eq.~(\ref{eq:Hs}). 
Because of its uniqueness, the steady state must be Gaussian. Indeed, both the local and global dynamics are Gaussian, so that any initial Gaussian state must converge to a Gaussian steady state.
If the initial Gaussian state is the vacuum, then $d(0)=0$; thus,  according to Eq.~(\ref{solution2}), $d(t)=0$ at all $t\geq 0$. Therefore, the local and global steady states denoted by $\rho^\infty_{\rm loc}$ and $\rho^\infty_{\rm glb}$ are Gaussian states with zero displacement vector and covariance matrices $C^\infty_{\rm loc}$ respectively $C^\infty_{\rm glb}$.

Due to Eq.~(\ref{eq:diffcovgen}), within the general Gaussian setting, the covariance matrix of the steady state $C^\infty$ must satisfy the 
following equality due to Eq.~(\ref{eq:diffcovgen})
\begin{equation}
\label{eq:covsteady}
\mathcal{M}\,C^{\infty}\,+\,C^{\infty}\,\mathcal{M}^\dag=-\mathcal{N},
\end{equation}
where $C^\infty$ is the covariance matrix of the steady state
and have the following form
\begin{equation}
\label{eq:StructorCSteadyLocal}
    C^{\infty}=\begin{pmatrix}
    C_1& C_2 \\
    C_2^\dag&C_1^{\top}.
    \end{pmatrix},\;\;C_1=C_1^\dag.
\end{equation}
As it is seen in Eqs.~(\ref{eq:MNLocal}), (\ref{eq:NLocal}), (\ref{eq:MNGlobal}), and (\ref{eq:NGlobal}), $\mathcal{M}$ and $\mathcal{N}$ have the following form
\begin{equation}
\mathcal{M}=\mathsf{M}\oplus\mathsf{M},\quad\mathcal{N}=\mathsf{N}\oplus\mathsf{N}.
\end{equation}
Therefore, to find $C^{\infty}$ we must solve Eq.~(\ref{eq:covsteady}) which yields the following equations for $C_1$ and $C_2$:
\begin{align}
\label{eq:c1General}
    &\mathsf{M}C_1+ C_1\mathsf{M}^\dag=-\mathsf{N},\\
    &\mathsf{M}C_2+ C_2\mathsf{M}=\bf{0}.
    \label{eq:c2steadyGeneral}
\end{align}
Because of the uniqueness of the steady state $C_2=\bf{0}$ is the only solution of Eq.~Eq.~(\ref{eq:c2steadyGeneral}). To solve Eq.~(\ref{eq:c1General}) for $C_1$ it is convenient to vectorise this equation: 
\begin{equation}
\label{eq:vecsteadyGeneral}
F\ket{C_1} = -\ket{\mathsf{N}},\quad F=(\mathsf{M}\otimes \mathbb{I}+\mathbb{I}\otimes \mathsf{M}^*),
\end{equation}
where $\ket{C_1}$, and $\ket{\mathsf{N}}$ are respectively the vectorized form of $C_1$, and $\mathsf{N}$. The formal solution of Eq.~(\ref{eq:vecsteadyGeneral}) is given by
\begin{equation}
\label{eq:ketc1General}
\ket{C_1} = -F^{-1}\ket{\mathsf{N}}.
\end{equation}
This formalism for finding the steady state is valid for both the local and global approach. Therefore, in both approaches, $C_2=0$. The possibility of deriving an analytical expression for $C_1$ depends on the possibility of inverting $F$ analytically. We will discuss it in further detail in the following subsections. 

\subsection{local approach}
\label{sec:Steadyloc}
In this subsection, we discuss how the steady state of the local approach can be formally derived. After deliberating on the complications for deriving a compact analytical expression for the covariance matrix of the steady state in the local approach, we follow the perturbation technique to derive the required covariance matrix up to the second order of $g$. 

To find the steady state, one must solve Eq.~(\ref{eq:ketc1General}) for $\ket{C_1}$ when $F$ is replace by $F_{\rm loc}$. The
analytical solution for $C_1$ is available when $J_\ell(\omega_0)=J_r(\omega_0)=J(\omega_0)$
and the two baths have the same temperature (for details see Appendix~\ref{app:LSSSpecialCase}). But, in a most general setting, it is not possible to find an analytical expression for the inverse of 
$F_{\rm loc}$. To find an analytical expression, we take the coupling constant $g$ as a perturbation parameter and rewrite $\mathsf{M}_{\rm loc}$ in Eq.~(\ref{eq:Mloc}):
\begin{equation}
    \mathsf{M}_{\rm loc}=\mathsf{M}_0+g\mathsf{M}_1,
\end{equation}
with $\mathsf{M}_0$ and $\mathsf{M}_1$
\begin{align}
    \mathsf{M}_0&=-i\omega_0\mathbb{I}-i\lambda^2H^{\rm (LS)}_{\rm loc}-\pi\lambda^2\mathsf{J}_{\rm loc},\\
    \mathsf{M}_1&=-i\begin{pmatrix}
        0&1&0\\
        1&0&1\\
        0&1&0\\
    \end{pmatrix}.
\end{align}
In Appendix~\ref{app:Perturbative steady state} it is shown how to derive $C_1$ to the second order of $g$:
\begin{equation}
    \label{eq:locsteadyg2}
    C_1=C^{(0)}+gC^{(1)}+g^2C^{(2)},
\end{equation}
where $C^{(0)}$, $C^{(1)}$ and $C^{(2)}$ are
\begin{align}
\label{eq:ketc_0_mat}
C^{(0)}&=\begin{pmatrix}
\tau_\ell(\omega_0)&0&0\cr
0&x&0\cr
0&0&\tau_r(\omega_0)
\end{pmatrix},\\
\label{eq:ketc_1_mat}
C^{(1)}&=-i\begin{pmatrix}
0&\frac{\tau_\ell(\omega_0)-x}{m_2}&0\cr
\frac{x-\tau_\ell(\omega_0)}{m_2^*}&0&\frac{x-\tau_r(\omega_0)}{m_8^*}\cr
0&\frac{\tau_r(\omega_0)-x}{m_8}&0\end{pmatrix}\ ,\\
\label{eq:ketc_2_mat}
C^{(2)}&=\begin{pmatrix}
\frac{\tau_\ell(\omega_0)-x}{|m_2|^2}&0&\frac{\tau_\ell(\omega_0)-x}{m_2m_3}+\frac{\tau_r(\omega_0)-x}{m_3m_8^*}\cr
0&0&0\cr
\frac{\tau_\ell(\omega_0)-x}{m_2^*m_3^*}+\frac{\tau_r(\omega_0)-x}{m_3^*m_8}&0&\frac{\tau_r(\omega_0)-x}{|m_8|^2}
\end{pmatrix}.
\end{align}
In these equations $\tau_\ell(\omega_0)$ and $\tau_r(\omega_0)$ were defined in Eq.~(\ref{eq:taualpha}), and
\begin{align}
    \label{eq:freeparameter}
    &x=\frac{|m_8|^2\Re(m_2)\tau_\ell(\omega_0)+|m_2|^2\Re(m_8)\tau_r(\omega_0)}{|m_8|^2\Re(m_2)+|m_2|^2\Re(m_8)},\\
    &m_2=-\lambda^2(iS_\ell(\omega_0)+\pi J_\ell(\omega_0)),\cr
    &m_8=-\lambda^2(iS_r(\omega_0)+\pi J_r(\omega_0)),\cr
    &m_3=m_2+m_8^*.\nonumber
\end{align}
\begin{remark}
\label{rem3}
In Appendix~\ref{app:Perturbative steady state} it is shown that the parameter $x$ is fixed by going to the second-order perturbation term.
In general, the zeroth-order contribution (Eq.~(\ref{eq:ketc_0_mat})) to the covariance matrix cannot be proportional to the identity. Therefore, the local steady state cannot be a thermal state unless the bath temperatures and couplings are the same at the two ends. In this case, $x=\tau_\ell(\omega_0)=\tau_r(\omega_0)$ and one retrieves  
the 
$g$-independent 
covariance matrix $ C^\infty_{\rm loc}=\tau_\ell(\omega_0)\mathbb{I}_6$ discussed in Appendix~\ref{app:LSSSpecialCase}. 
\end{remark}

\subsection{global approach}
\label{sec:Steadyglb}

In this subsection, we discuss how to find the steady state of the master equation in the global approach. We discuss two different techniques and derive the explicit form of the covariance matrix of the steady state. 

The first technique is the general one discussed at the beginning of \S\ref{Sec:SteadyState}. Following that general discussion, in Eq.~(\ref{eq:ketc1General}), we replace $F$ and $\ket{\mathsf{N}}$ with $F_{\rm glb}$ and $\ket{\widetilde{\mathsf{N}}_{\rm glb}}$ to find $\ket{\widetilde{C}_1}$. $F_{\rm glb}$ is given in Eq.~(\ref{eq:vecsteadyGeneral}), when $\mathsf{M}$ is replaced by $\widetilde{\mathsf{M}}_{\rm glb}$ in 
Eq.~(\ref{eq:Mglobal1}) and $\ket{\widetilde{\mathsf{N}}_{\rm glb}}$ is the vectorized form of $\widetilde{\mathsf{N}}_{\rm glb}$ given in Eq.~(\ref{eq:Nglobal1}). 
$\ket{\widetilde{C}_1}$ is the vectorized form of the first block of the global steady state in $\hat{\xi}_c$ basis. One can return to the representation in $\hat{\xi}_a$ by the similarity transformation $T^\dagger$ (see Eq.~(\ref{eq:basis})). Because of the diagonal form of $\mathsf{M}_{\rm glb}$, $F_{\rm glb}$ is diagonal and its inverse is computable. Therefore, unlike the local approach, in the global approach we have the explicit form of 
$\widetilde{C}_1$
\begin{equation}
    (\widetilde{C}_1)_{ii}=\frac{\sum_{\alpha=l,r}J_\alpha(\epsilon_i)(2\bar{n}_{\alpha}(\epsilon_i)+1)}{\sum_{\alpha=l,r}J_\alpha(\epsilon_i)}.
\end{equation}

In the case of the same spectral density for two baths, $J_\ell(\epsilon_i)=J_r(\epsilon_i)=J(\epsilon_i)$, the covariance matrix of the global steady state in $\hat{\xi}_c$ basis is
\begin{equation}
\label{eq:C1globalGeneral}
   \widetilde{C}^\infty_{\rm glb}=\widetilde{C}_1\oplus \widetilde{C}_1^\top,\quad \widetilde{C}_1=\operatorname{diag}\big(\tau(\epsilon_1),\tau(\epsilon_2),\tau(\epsilon_3)\big), 
\end{equation}
with $\tau(\epsilon_i)$
\begin{equation}
\label{eq:tauepsiloni}
    \tau(\epsilon_i)=\sum_{\alpha=\ell,r}\bar{n}_\alpha(\epsilon_i)+1.
\end{equation}

\begin{remark}
\label{rem4}
We emphasize that when the inter-oscillator coupling is removed ($g=0$), Eq.~(\ref{eq:H_sglb}) yeilds $\epsilon_1=\epsilon_2=\epsilon_3=\omega_0$ . Therefore, at $g=0$ the covariance matrix is proportional to the identity matrix. Recalling Remark~\ref{rem3}, it reveals an important distinction: the steady state of the global approach differs fundamentally from that of the local approach at $g=0$. The two steady states derived from both approaches coincide at $g=0$ if $T_\ell=T_r$. In this case 
\begin{equation}
    C^\infty_{\rm loc}=C^\infty_{\rm glb}=\tau_\ell(\omega_0)\mathbb{I}_6.
\end{equation}
\end{remark}
One can compare the above algebraic derivation with the following argument, based on Remark~\ref{rem2}. According to Eq.~(\ref{3factors}), the system's dynamic is generated by three independent generators in $\hat{\xi}_c$ basis. Therefore, the steady state is the product of three Gaussian states $\rho_j^{\infty}$, which are the solution of $\mathcal{L}_j[\rho_j^\infty]=0$ for $j=1,2,3$ and $\mathcal{L}_j$ as defined in Eq.~(\ref{3factors}). Therefore,
\begin{equation}
\label{steadyglb1}
\rho^\infty_{\rm glb}=\prod_{j=1}^3\rho^\infty_j,\quad \rho^\infty_j=\Big(1-{\rm e}^{-\gamma_j}\Big)\,{\rm e}^{-\gamma_j c^\dag_j\,c_j},
\end{equation}
where$\gamma_j=\log \frac{b_j}{a_j}$
with
\begin{equation}
a_j:=\sum_{\alpha=l,r}J_{\alpha}(\epsilon_j)\bar{n}_{\alpha}(\epsilon_j)\ ,\ b_j:=a_j+\sum_{\alpha=l,r}J_{\alpha}(\epsilon_j).
\end{equation}
Therefore, the steady state is a Gaussian thermal state with no displacement vector and diagonal covariance matrix 
$ \widetilde{C}^{\infty}_{\rm glb}=\widetilde{C}_1\oplus\widetilde{C}_1^{\top}$ in Eq.~(\ref{eq:C1globalGeneral}).

\section{Local approach versus global approach}
\label{sec:comparing-steady-state}

Equipped with the local and global dynamics and with the corresponding steady states, we can now proceed and assess the validity of the two approaches as approximation of the exact dynamics of the open quantum harmonic chain.
Firstly, we shall compare the steady states and then their time-evolution under the exact dynamics.

In \S\ref{Sec:SteadyState}, we have seen that in both the local and global approaches there is a unique steady state and that the local steady state $\rho^\infty_{\rm loc}$
does not emerge from the global one, $\rho^\infty_{\rm glb}$, by switching off the chain internal interactions. In order to further the comparison of the two steady states and we move away from $g=0$.
As a measure of their difference we use the fidelity which is computable in terms of covariance matrices and displacement vectors (see Appendix~\ref{app:fiddelity}).

\begin{figure}
    \centering
    \includegraphics[width=1\linewidth]{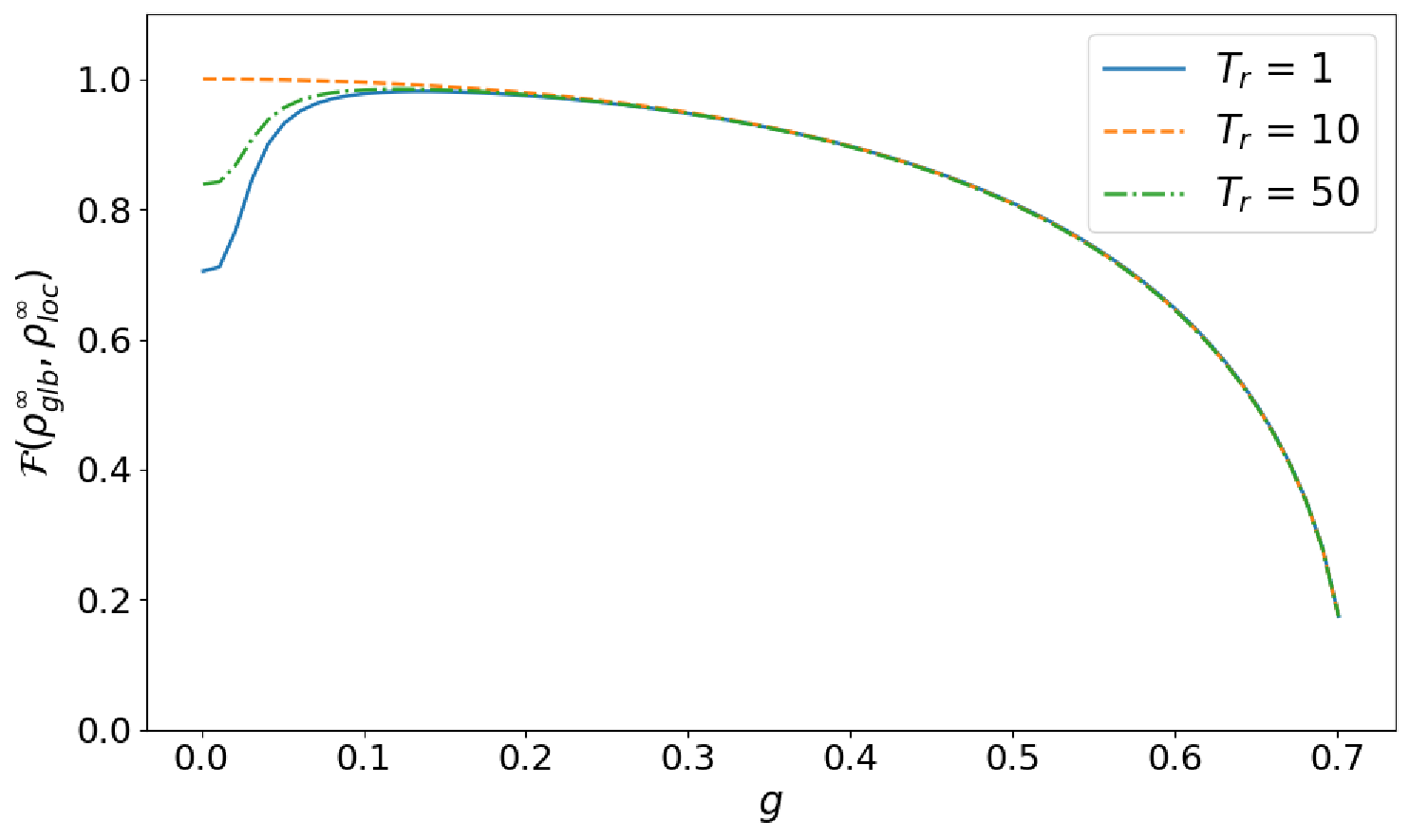}
    \caption{Fidelity between $\rho_{\rm loc}^\infty$ and $\rho_{\rm glb}^\infty$ versus $g$ for $\lambda=0.1$, $\omega_0=1$, $\omega_c=3$ and $T_\ell=10$. 
    The blue-solid curve, orange-dashed curve and green-dash-dotted curve are respectively for $T_r=1,10,50$.}
    \label{fig:fiddelity-versus-g}
\end{figure}

In the special case of equal bath temperatures, ($T_\ell=T_r$), the fidelity between the two steady states admits an analytical solution:
\begin{align}
\label{eq:fidelity-analytic}
    &\mathcal{F}^4(\rho^\infty_{\rm loc}, \rho^\infty_{\rm glb})\cr
&=\prod_{i=1}^3\frac{\big(\sqrt{(4\tau^2(\epsilon_2)-1)(4\tau^2(\epsilon_i)-1)}+4\tau(\epsilon_2)\tau(\epsilon_i)+1\big)^2}{4(\tau(\epsilon_2)+\tau(\epsilon_i))^4},\cr
\end{align}
where $\tau(\epsilon_i)$ is defined in Eq.~(\ref{eq:tauepsiloni}). The orange-dashed curve in Fig.~(\ref{fig:fiddelity-versus-g}), shows $\mathcal{F}(\rho^\infty_{\rm loc}, \rho^\infty_{\rm glb})$ in Eq.~(\ref{eq:fidelity-analytic}) versus $g$, for $T_\ell=T_r=10$. 

When $T_r\neq T_\ell=10$, the fidelity is computed numerically for $T_r=1$ and $T_r=50$; its behavior versus $g$ is provided in Fig.~\ref{fig:fiddelity-versus-g}. Furthermore, for all numerical results in this section, we set $\lambda=0.1$, $\omega_0=1$, 
$\omega_c=3$.
Fig.~\ref{fig:fiddelity-versus-g} shows the existence of a threshold value of the coupling constant, denoted by $g^*$. 
For $g<g^*$, by increasing $g$, $\mathcal{F}(\rho_{\rm loc}^\infty,\rho_{\rm glb}^\infty)$ increases. Conversely, for $g>g^*$, increasing $g$ results in decreasing $\mathcal{F}(\rho_{\rm loc}^\infty,\rho_{\rm glb}^\infty)$. When $T_\ell=T_r$ the threshold value $g^*$ vanishes. As already discussed in Remark~\ref{rem4}, at $g=0$ and equal temperatures, the two steady states coincide and the fidelity is $1$ as shown by the orange-dashed curve in Fig.~\ref{fig:fiddelity-versus-g}.
When the two baths temperatures are identical $\mathcal{F}(\rho_{\rm loc}^\infty,\rho_{\rm glb}^\infty)$ is a decreasing function of $g$. But when $T_\ell\neq T_r$, the inter-oscillator coupling $g$ must be larger than the threshold value $g^*$ to put into evidence  the difference between the local and global steady states.

Fig.~\ref{fig:fiddelity-versus-g} reveals that, for small values of $g$, the two bath temperatures affect the fidelity between the two steady states. 
On the other hand,  for large values of $g$, the fidelities for different values of $T_r$ coincide and decrease so that
the two steady states deviate from each other no matter what the baths temperatures are. 

 \begin{figure}
    \begin{subfigure}[H]{0.5\textwidth}   \centering
    \includegraphics[width=\textwidth]{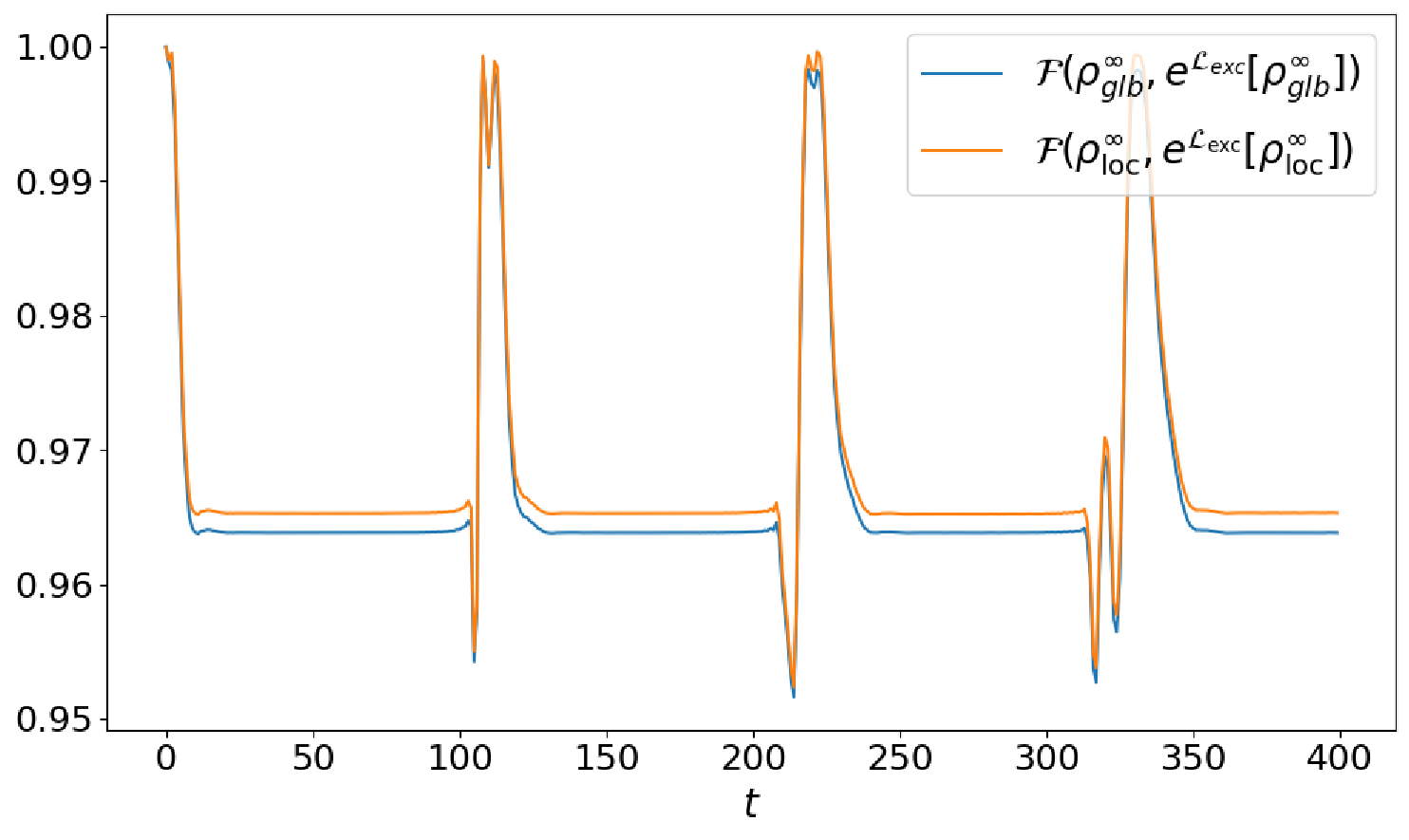}
            \caption{$g=0.01$}
            \label{fig:g=0.01}
    \end{subfigure}%
   
    \begin{subfigure}[H]{0.5\textwidth}
            \centering
     \includegraphics[width=\textwidth]{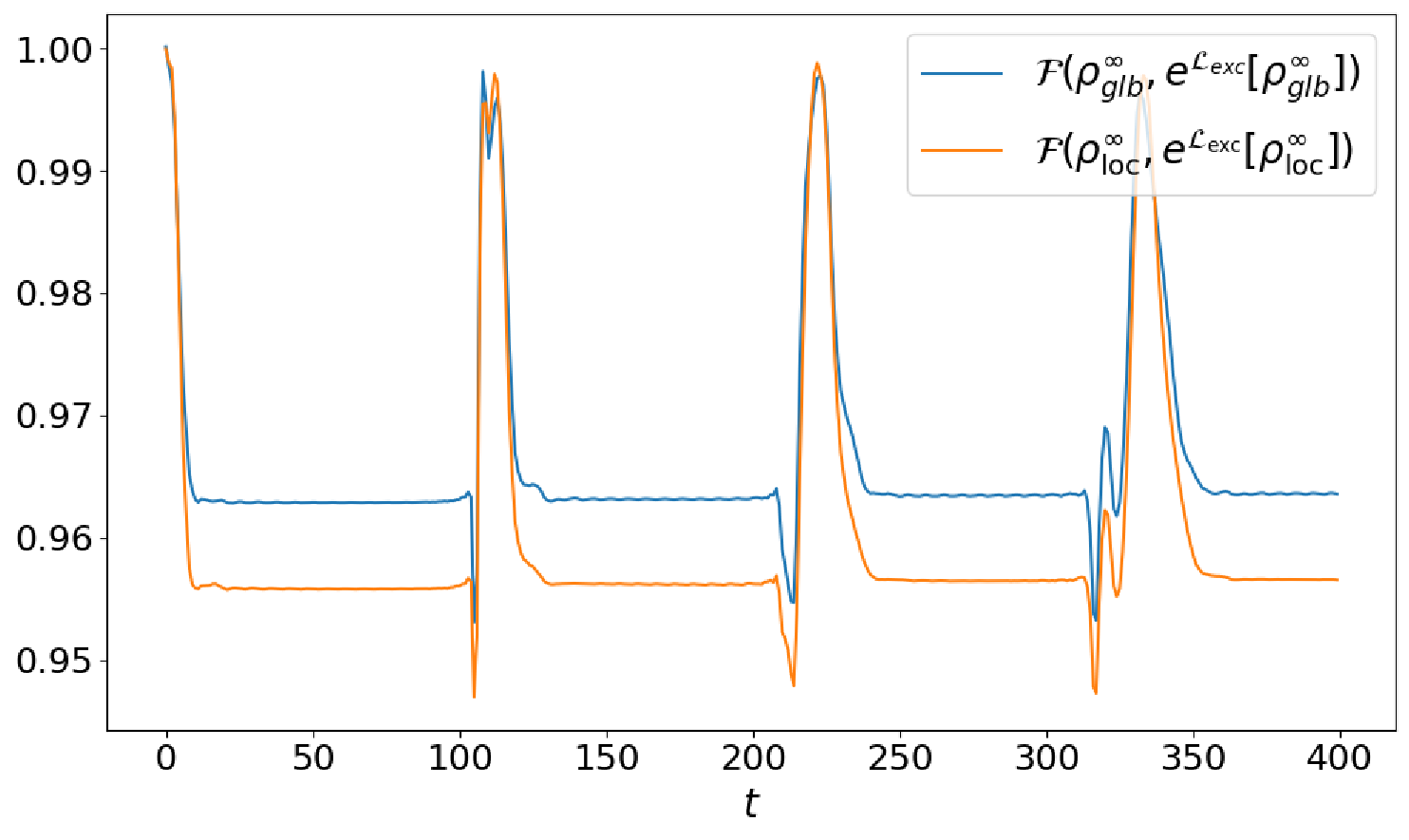}
            \caption{$g=0.1$}
            \label{fig:g=0.1}
    \end{subfigure}
    
    \begin{subfigure}[H]{0.5\textwidth}
            \centering
      \includegraphics[width=\textwidth]{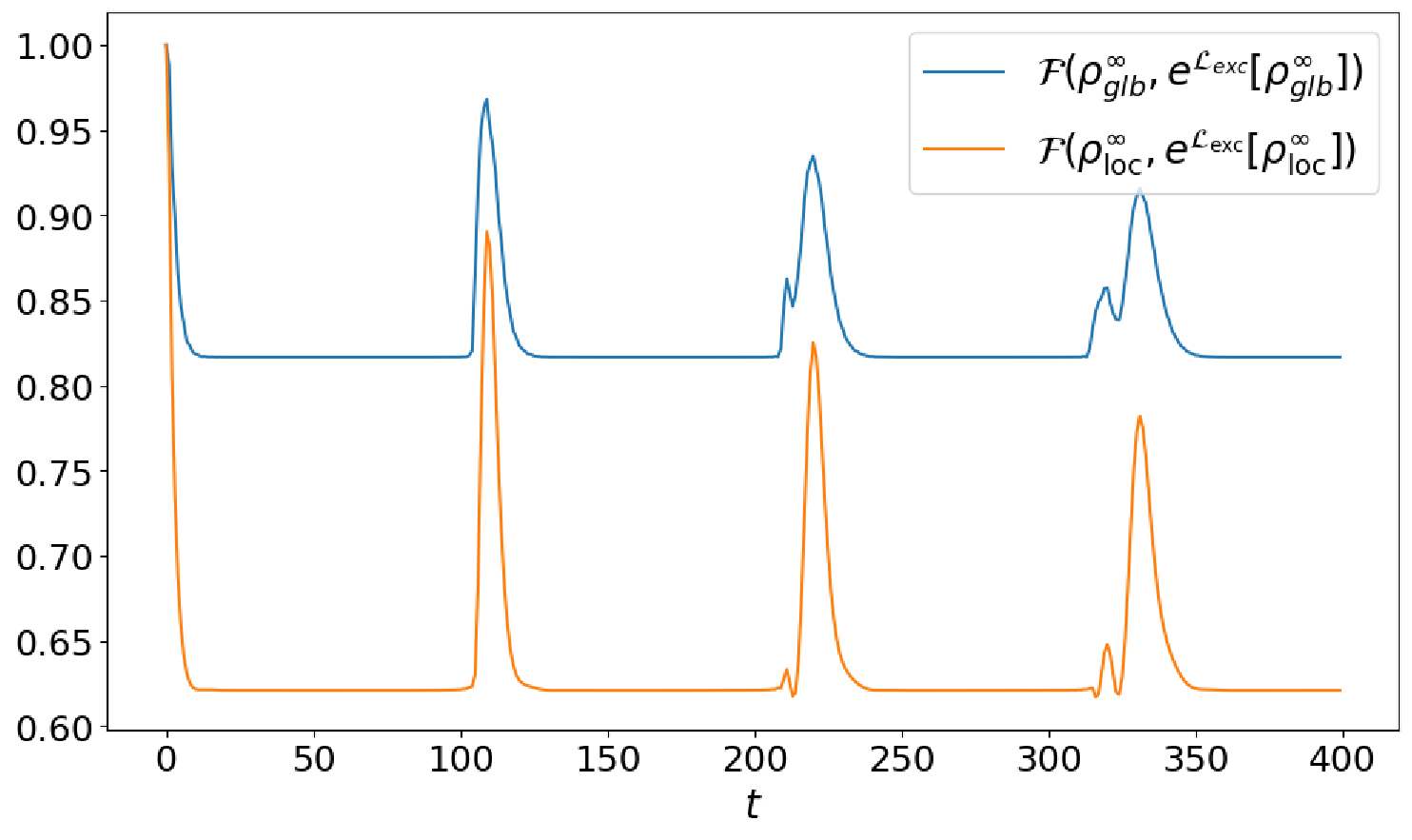}
            \caption{$g=0.6$}
            \label{fig:g=0.6}
    \end{subfigure}
    
    \caption{Fidelity between $\rho_{\rm loc}^\infty$ and the evolved $\rho_{\rm loc}^\infty$ under the evolution of the exact approach i.e $\mathcal{F}(\rho_{\rm loc}^\infty,e^{t\mathcal{L}_{\rm exc}}[\rho_{\rm loc}^\infty])$ (orange curve) and 
    fidelity between $\rho_{\rm glb}^\infty$ and the evolved $\rho_{\rm glb}^\infty$ under the evolution of the exact approach i.e $\mathcal{F}(\rho_{\rm glb}^\infty,e^{t\mathcal{L}_{\rm exc}}[\rho_{\rm glb}^\infty])$ (blue curve) versus time $t$. From top to bottom $g=0.01,0.1,0.6$. In all sub figures $\lambda=0.1$, $\omega_0=1$, $\omega_c=3$, $T_\ell=10$, $T_r=8$ and $M=M'=50$.}
    \label{fig:evolution-loc-glb}
\end{figure}

Up to here, our analysis has established a clear comparison between the steady states obtained from the local and global approaches. 
To evaluate the validity of each approach, we must compare their predictions against the exact solution of the open harmonic chain. 
The exact evolution of the chain and its environment is unitary, governed by the total Hamiltonian. As a result, all system properties exhibit oscillatory dynamics. However, when the baths contain a sufficiently large number of modes to effectively mimic infinite reservoirs, we expect that an almost stationary behavior is reached before periodicity gets in. This periodic stationarity with periods becoming larger with increasing numbers of environmental modes is a distinctive mark that ensures the validity of both the numerical computation of the exact unitary dynamics and its local and global irreversible reduced dynamics approximations. 

To asses which reduced dynamics is preferable and in which regime, we now study the behavior of the steady states under the exact evolution. The idea is that the more accurate description is provided by the steady state which less deviate from itself when evolved in time by the exact evolution.
To inspect this behavior, we numerically compute $\mathcal{F}(\rho_{\rm loc}^\infty,e^{t\mathcal{L}_{\rm exc}}[\rho_{\rm loc}^\infty])$ and $\mathcal{F}(\rho_{\rm glb}^\infty,e^{t\mathcal{L}_{\rm exc}}[\rho_{\rm glb}^\infty])$.
In these expressions, $e^{t\mathcal{L}_{\rm exc}}[\rho]$ denotes
\begin{equation}
    e^{\mathcal{L}_{\rm exc}}[\rho]= \operatorname{Tr}_B(e^{-it\hat{H}_{\rm tot}}(\rho\otimes\rho_{\beta_\ell}\otimes\rho_{\beta_r})e^{it\hat{H}_{\rm tot}}),
\end{equation}
where $\rho_{\beta_\ell}$, respectively $\rho_{\beta_r}$, stand for the thermal states of the the left and right bath with $T_\ell$ and $T_r$.
Figure~\ref{fig:evolution-loc-glb} shows $\mathcal{F}(\rho_{\rm loc}^\infty,e^{t\mathcal{L}_{\rm exc}}[\rho_{\rm loc}^\infty])$ (orange curve)
and $\mathcal{F}(\rho_{\rm glb}^\infty,e^{t\mathcal{L}_{\rm exc}}[\rho_{\rm glb}^\infty])$ (blue curve) versus $t$ for different values of $g$. From top to bottom $g=0.01,0.1,0.6$
and in all sub-figures, $T_\ell=10$, $T_r=8$.
Beside clearly showing the periodic quasi-stationary sought after, Fig.~\ref{fig:evolution-loc-glb} shows that, depending on the inter-oscillator coupling constant $g$, the more accurate approach is not necessarily the global one. In fact, there is a critical value for the inter-oscillator coupling strength denoted by $g_c$. For $g<g_c$: 
\begin{equation}
\label{locbetter}
    \mathcal{F}(\rho_{\rm loc}^\infty,e^{t\mathcal{L}_{\rm exc}}[\rho_{\rm loc}^\infty])>\mathcal{F}(\rho_{\rm glb}^\infty,e^{t\mathcal{L}_{\rm exc}}[\rho_{\rm glb}^\infty]) ,
\end{equation}
and for $g>g_c$
\begin{equation}
\label{glbetter}
\mathcal{F}(\rho_{\rm loc}^\infty,e^{t\mathcal{L}_{\rm exc}}[\rho_{\rm loc}^\infty])<\mathcal{F}(\rho_{\rm glb}^\infty,e^{t\mathcal{L}_{\rm exc}}[\rho_{\rm glb}^\infty]) .
\end{equation}
This behavior puts into evidence that when $g<g_c$, the local steady state deviates less from itself under the exact dynamics with respect to what becomes of the global steady state. Looking closely at the graphs, 
we see that for $g=0.01$ although the orange and blue curves are very close to each other, none the less the local approach outperforms the global one. Instead, increasing $g$ beyond a critical $g_c$, the global approach is to be preferred to the local one (see sub-figures in Fig.~\ref{fig:evolution-loc-glb} with $g=0.1$ and $g=0.6$).

\begin{figure}
    \centering
    \includegraphics[width=1\linewidth]{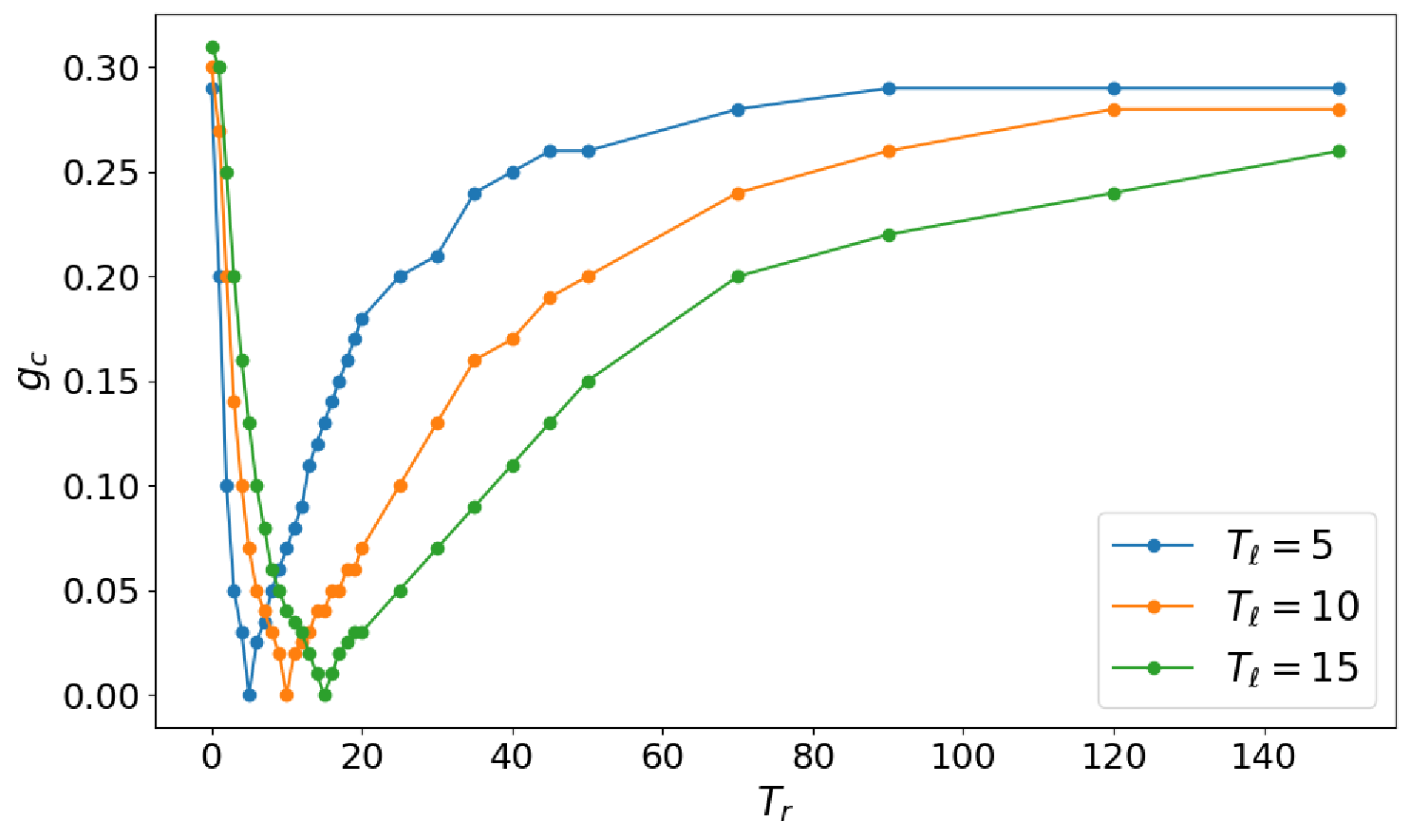}
    \caption{Here, we plot $g_c$ versus $T_r$. The blue, orange and green lines are for $T_\ell=5,10,20$, respectively.  we set $\lambda=0.1$, $\omega_0=1$, $\omega_c=3$, $M=M'=50$ and $t=50$.}
    \label{fig:g-criticalversusT_r}
\end{figure}

In conclusion, when
$g<g_c$, the quasi-stationary state associated with the exact reversible dynamics is more closely reproduced by the local steady state. On the other hand, 
when $g>g_c$, it is the global steady state which better reproduces the exact state of affairs.
It is thus clear that there exists a critical value of the harmonic chain internal couplings,  $g_c$, that sorts out the local and the global approaches to the derivation of a suitable reduced dynamics from the open harmonic chain. Such a threshold value depends on the two bath's temperatures.  In fact,
Fig.~\ref{fig:g-criticalversusT_r} presents the behavior of $g_c$ versus $T_r$ when $T_\ell=5$ (blue curve), $T_\ell=10$ (orange curve) and $T_\ell=15$ (green curve). When $T_\ell=T_r$, $g_c=0$. This means that when $T_\ell=T_r$, the global approach is better than the local approach for couplings $g$. When $\abs{T_\ell-T_r}$ increases, $g_c$ increases.
\section{conclusion}
\label{sec:Conclusion}

In this contribution, we have investigated the dynamics of a three-harmonic oscillator chain, where the first and third one of them interact with two different thermal baths at different temperatures. We analyze the reduced dynamics of the open harmonic chain using three different approaches: local, global, and exact.
As a first step, we derived the master equations that lead to the local and global reduced dynamics. We then found the unique Gaussian steady states in both approaches, analytically in the case of the global steady state, perturbatively for the local one.
Subsequently, to assess their vicinity to the actual, that is to the non approximated,  behavior of the open harmonic chain, we examined the fate of these steady states under the exact time-evolution. Specifically, we compared the behavior in time of the fidelity of the local and global steady states with respect to their time-evolved states under the exact dynamics.
Besides setting a useful Gaussian framework for addressing the issue of  which approach is to be preferred under which physical circumstances, 
the main result of our analysis is having given evidence to the existence of a critical inter-oscillator coupling strength $g_c$.
For $g<g_c$, the local approach provides a better approximation, while for $g>g_c$, the global approach becomes more accurate. Furthermore, we found that the critical coupling $g_c$ depends on the temperature difference between the two baths. As this temperature difference decreases, the range of $g$ for which the global approach is superior expands.
In summary, we have shown that 1) the reduced dynamics of the local approach cannot be retrieved from the global one by simply letting the inter-couplings vanish, 2) that the 
simplified local approach can, for sufficiently small couplings outperform the orthodox global approach. This, in our opinion, indicates the necessity of further 
and more refined studies  of how the the weak-coupling-limit techniques  have to be applied in the case of many-body open quantum systems.

\begin{acknowledgments}
	F.B. acknowledges financial support from PNRR MUR project PE0000023-NQSTI. L.M. and M.B acknowledge financial support from the Iran National Science Foundation (INSF) under Project No. 4022322. L. M. acknowledges support from the ICTP through the Associates Pro- gramme (2019-2024). M. B. acknowledges financial support from International Center for Theoretical Physics (ICTP) under the Sandwich Training Education Program (STEP). 

\end{acknowledgments}

\appendix
\section{Example}
\label{App:Exam}
In this appendix, we discuss a paradigmatic example. let us consider two non-interacting harmonic oscillators with frequencies $\omega_i$ in thermal equilibrium at inverse temperatures 
$\beta_{i}=\frac{1}{T_{i}}$, $i=1,2$:
\begin{equation}
\label{2modes0}
\rho_{12}=\Big(1-{\rm e}^{-\beta_1\omega_1}\Big)
\Big(1-{\rm e}^{-\beta_2\omega_2}\Big)\,{\rm e}^{-\beta_1\omega_1a^\dag_1 a_1-\beta_2\omega_2a^\dag_2a_2}\ .
\end{equation}
We recall that we work in the natural units where Boltzman constant
$\kappa_B = 1$.
Introducing the mean occupation numbers of mode $i$ with frequency $\omega_i$:
\begin{equation}
\label{2modes1}
\bar{n}(\omega_i)={\rm Tr}\Big(\rho_{12}\,a^\dag_{i}a_{i}\Big)=\frac{1}{{\rm e}^{\beta_{i}\omega_{i}}-1},
\end{equation}
the covariance matrix is diagonal, which signals the lack of correlations between the two oscillators:
\begin{equation}
\label{2modes2}
C=\begin{pmatrix}
    \mathcal{C}_1& 0\\
    0& \mathcal{C}_1
    \end{pmatrix}\ ,\
    \mathcal{C}_1=\operatorname{diag}(2\bar{n}(\omega_1)+1,2\bar{n}(\omega_2)+1)\ .
\end{equation}
For oscillators with the same frequencies and at equal temperatures $\bar{n}(\omega_1)=\bar{n}(\omega_2)=\bar{n}$, $C=(2n+1)\mathbb{I}_{2N}$. Furthermore, for the vacuum state which is reached 
when $T\to0$ ( $\beta\to+\infty$), $n\to0$ and $C=\mathbb{I}_{2N}$.

\section{Master equation for the covariance matrix and the displacement vector}
\label{sec:MasterCD}
In this appendix, we sketch the derivation of the equations of motions of the covariance matrix and displacement vector of any initial Gaussian state that evolve 
according to the master equation in Eq.~({\ref{eq:general masterequation}})
The adjoint or dual dynamics in the Heisenberg picture corresponding to the Schr\"odinger picture dynamics in Eq.~(\ref{eq:general masterequation}), reads:
\begin{equation} 
\label{eq:adjointGen}
\dot{O}=\mathcal{L}^{\rm adj}[O],
\end{equation}
where
\begin{align}
\label{eq:mas}
\mathcal{L}^{\rm adj}[\bullet]&=i[\hat{H},\bullet]+\lambda^2\mathcal{D}^{\rm adj}[\bullet],
\end{align}
is the adjoint of the generator $\mathcal{L}$ in (\ref{eq:general masterequation}), with
\begin{equation}
\label{eq:adjdiss}
    \mathcal{D}^{\rm adj}[\bullet]=\sum_{k}\alpha_k(\xi_k\bullet\xi_k^\dag-\frac{1}{2}\{\xi_k\xi_k^\dag,\bullet\}).
\end{equation}
By taking time derivative of the covariance matrix and displacement vector in Eq.~(\ref{eq:covaadag}) and using Eq.~(\ref{eq:adjointGen}) we have:
\begin{align}
\label{eq:dotcd}
\dot{C}_{ij}&=\expval{\mathcal{L}^{\rm adj}[\xi_i\xi_j^\dag+\xi_j^\dag\xi_i]}\cr
&-2(\expval{\mathcal{L}^{\rm adj}[\xi_i]}\expval{\xi_j^\dag}+\expval{\xi_i}\expval{\mathcal{L}^{\rm adj}[\xi_j^\dag]}),\cr
\dot{d}_i&=\expval{\mathcal{L}^{\rm adj}[\xi_i]}.
\end{align}
Therefore, to derive the master equation for the covariance and the displacement vector, we need to compute $\mathcal{L}^{\rm adj}[\xi_i]$ and $\mathcal{L}^{\rm adj}[\xi_i\xi_j^\dag+\xi_j^\dag\xi_i]$. 

For the quadratic Hamiltonian in Eq.~(\ref{eq:quadratic-Hamiltonian}) and using the commutation relation of bosonic operators in Eq.~(\ref{eq:Omega}), and the adjoint of $\mathcal{D}[\bullet]$ in Eq.~(\ref{eq:adjdiss}) we have
\begin{align}
\label{eq:adjGenXi}
\mathcal{L}^{\rm adj}[\hat{\xi}]&= iW\hat{\xi}-\lambda^2 D\mathbb{X}_N\Omega\hat{\xi},\\
\mathcal{L}^{\rm adj}[\xi_i\xi_j^\dag+\xi_j^\dag \xi_i] &= i\sum_nW_{jn}(\xi_i\xi_n^\dag+\xi_n^\dag\xi_i)\cr
&+ \frac{\lambda^2}{2}\sum_{n}(D\mathbb{Z}_N+\mathbb{X}_ND\Omega)_{jn}(\xi_n^\dag\xi_i+\xi_i\xi_n^\dag)\cr
&+\lambda^2(D-\Omega D\Omega)_{ij}
\label{eq:adjGenXiiXij}
\end{align}
where $W$ and $\mathbb{X}_N$ and $D$ are
\begin{align*}
    W&=-\Omega(\mathsf{H}\mathbb{X}_N+\mathbb{X}_N\mathsf{H}),\\
    \mathbb{X}_N &=\begin{pmatrix}
        {\bf{0}}&\mathbb{I}_N\\
        \mathbb{I}_N&{\bf{0}}\\
    \end{pmatrix},\\
    D_{ij}&=\alpha_i \delta_{ij}. 
\end{align*}
Therefore, by considering Eq.~(\ref{eq:dotcd}), Eq.~(\ref{eq:adjGenXi}) and Eq.~(\ref{eq:adjGenXiiXij}) the differential equation of the covariance matrix and the displacement vector are given by
\begin{align*}
\dot{C}(t)&=\mathcal{M}C(t)+C(t)\mathcal{M}^\dag+\mathcal{N},\cr
\dot{d}(t)&=\mathcal{M}d(t),
\end{align*}
where $\mathcal{M}$ and $\mathcal{N}$ are
\begin{align*}
    \mathcal{M}&=i W+\frac{\lambda^2}{2}(D\mathbb{Z}_N+\mathbb{X}_ND\Omega),\cr
   \mathcal{N}&=\lambda^2(D-\Omega D\Omega).
\end{align*}
The solution of the time derivative of the covariance matrix and displacement vector has been written in Eq.~(\ref{solution1}) and Eq.~(\ref{solution2}).

\section{The steady state of the local approach}
In this appendix, we discuss the steady state of the local approach. In App.~\ref{app:LSSSpecialCase}, we obtain the exact steady state solution for the special case of equal bath temperatures. In App.~\ref{app:Perturbative steady state}, we present the perturbative solution for the general bath temperatures, retaining terms up to the second order of the inter-oscillator coupling strength $g$.

\subsection{A special case}\label{app:LSSSpecialCase}
Here we derive the covariance matrix of the steady state for a special case: $J_{\ell}(\omega_0)=J_{r}(\omega_0)=: J(\omega_0)$ and the two bath temperatures are equal ($\bar{n}_\ell(\omega_0)=\bar{n}_r(\omega_0)=:\bar{n}(\omega_0)$). 
In such a case, due to the symmetry, we expect that the covariance matrix remains invariant under the change of labels $1$ and $3$. That gives the following form for the covariance matrix:
\begin{equation}
\label{eq:steadyc1loc}
    C_1=\begin{pmatrix}
        \alpha &\beta&\gamma\\
        \bar{\beta}&\delta&\beta\\
        \bar{\gamma}&\bar{\beta}&\alpha
    \end{pmatrix}.
\end{equation}
If we solve Eq.~(\ref{eq:c1General}) for $C_1$ in Eq.~(\ref{eq:steadyc1loc}), 
when $\mathsf{M}$ and $\mathsf{N}$ are replaced by $\mathsf{M}_{\rm loc}$ and $\mathsf{N}_{\rm loc}$, 
we get $C_1=\tau_\ell(\omega_0)\mathbb{I}_3$. Therefore, the covariance matrix of the unique steady state is given by
\begin{equation}
\label{eq:steady-local}
    C^\infty_{\rm loc}=\tau_\ell(\omega_0)\mathbb{I}_6\;,\;\; \;\tau_\ell(\omega_0)=2\bar{n}_\ell(\omega_0)+1.
\end{equation}
It implies that when the coupling to the left and right baths are symmetric and both have the same temperature, the steady state is a tensor product of the thermal states of each mode. 
\subsection{Perturbative steady state}
\label{app:Perturbative steady state}
Here we drive the local steady state with the perturbative approach up to the second order of the inter-oscillator coupling strength $g$.

By expanding $\ket{C_1}$, which is the vectorized form of the first block of the local steady state, in powers of $g$
\begin{equation}
    \ket{C_1}=\sum_{n=0}^\infty g^n\ket{C^{(n)}},
\end{equation}
and using Eq.~(\ref{eq:ketc1General}) when $F$ and $\ket{\mathsf{N}}$ are replaced by $F_{\rm loc}$ and $\ket{\mathsf{N}_{\rm loc}}$, we obtain the following recursive equations 
\begin{align}
\label{eq:ketC0}
    \ket{C^{(0)}}&=-\mathfrak{M}_0^{-1}\ket{\mathsf{N}_{\rm loc}},\\
     \ket{C^{(n+1)}}&=-\mathfrak{M}_0^{-1}\mathfrak{M}_1\ket{C^{(n)}},
     \label{eq:recursive}
\end{align}
where $\mathfrak{M}_n$ is
\begin{equation}
    \mathfrak{M}_n=\mathsf{M}_n\otimes\mathbb{I}+\mathbb{I}\otimes\mathsf{M}_n^*,\quad {\rm for}\quad n=0,1.
\end{equation}
To find $\ket{C^{(n+1)}}$ by using the recursive relation in Eq.~(\ref{eq:recursive}), we need to invert $\mathfrak{M}_0$ which has the following explicit form: 
\begin{equation}
\label{eq:M0}
   \mathfrak{M}_0=\operatorname{Diag}(m_1,m_2,...,m_9),
\end{equation}
with $m_i$
\begin{align}
    &m_1=-2\pi\lambda^2J_\ell(\omega_0),\cr
    &m_2=m_4^*=-\lambda^2(iS_\ell(\omega_0)+\pi J_\ell(\omega_0)),\cr
    &m_5=0,\cr
    &m_8=m_6^*=-\lambda^2(iS_r(\omega_0)+\pi J_r(\omega_0)),\cr
    &m_3=m_7^*=m_2+m_6,\cr
    &m_9=-2\pi\lambda^2J_r(\omega_0).
\end{align}
Because $\operatorname{det}(\mathfrak{M}_0)=0$, it is not invertible, unless we restrict it to a subspace orthogonal to its kernel. The latter,
from Eq.~(\ref{eq:M0}) is spanned by
\begin{equation}
\operatorname{Ker}({\mathfrak{M}_0})=\operatorname{Span}\{    \ket{\mathcal{K}}=(0,0,0,0,1,0,0,0,0)^\top\},
\end{equation}
On the other hand, from Eq.~(\ref{eq:Nloc}) we construct $\ket{\mathsf{N}_{\rm loc}}$ as follows:
\begin{align}
   \ket{\mathsf{N}_{\rm loc}}=2\pi\lambda^2&\big(J_\ell(\omega_0)\tau_\ell(\omega_0),0,0,0,0\cr
&,0,0,0,J_r(\omega_0)\tau_r(\omega_0)\big)^\top,
\end{align}
where $\tau_\alpha(\omega_0)$ has been defined in Eq.~(\ref{eq:taualpha}).
It is clear that $\ket{\mathsf{N_{\rm loc}}}\notin\operatorname{Ker}(\mathfrak{M}_0)$ therefore the right hand side of Eq.~(\ref{eq:ketC0}) is well-defined. By solving $\mathfrak{M}_0\ket{C^{(0)}}=-\ket{\mathsf{N_{\rm loc}}}$ we find the general form of $\ket{C^{(0)}}$ as follows
\begin{equation}
\label{eq:ketc_0}
    \ket{C^{(0)}}=(\tau_\ell(\omega_0),0,0,0,x,0,0,0,\tau_r(\omega_0))^\top.
\end{equation}
Here, $x$ is an unknown parameter not determined by considering the zero order of perturbation. Because $\mathfrak{M}_1\ket{C^{(0)}}\notin\operatorname{Ker}{\mathfrak{M}_0}$, the recursive relation in Eq.~(\ref{eq:recursive}) determines
$\ket{C^{(1)}}$:
\begin{align}
\label{eq:ketc_1}
    \ket{C^{(1)}}=-i(&0,\frac{\tau_\ell(\omega_0)-x}{m_2},0,\frac{x-\tau_\ell(\omega_0)}{m_4},0,\cr
    &\frac{x-\tau_r(\omega_0)}{m_6},0,\frac{\tau_r(\omega_0)-x}{m_8},0)^\top.
\end{align}
To determine the unknown parameter $x$ we need to go to higher orders of perturbation. For computing $\ket{C^{(2)}}$ we must again use the recursive relation 
in Eq.~(\ref{eq:recursive}). But  $\mathfrak{M}_1\ket{C^{(1)}}\in\operatorname{Ker}(\mathfrak{M}_0)$. 
To guarantee the possibility of going to higher orders of perturbation, we determine $x$ such that $\mathfrak{M}_1\ket{C^{(1)}}\notin\operatorname{Ker}(\mathfrak{M}_0)$ which is equivalent to solve $\bra{\mathcal{K}}{\mathfrak{M}_1 \ket{C^{(1)}}}=0$ for $x$
\begin{equation*}
    x=\frac{|m_8|^2\Re(m_2)\tau_\ell(\omega_0)+|m_2|^2\Re(m_8)\tau_r(\omega_0)}{|m_8|^2\Re(m_2)+|m_2|^2\Re(m_8)}.
\end{equation*}
With $x$ in Eq.~(\ref{eq:freeparameter}), $\ket{C^{(2)}}$ is computed by using the recursive relation in Eq.~(\ref{eq:recursive}).
\begin{align}
    \label{eq:ketc_2}\ket{C^{(2)}}&=\big(\frac{\tau_\ell(\omega_0)-x}{|m_2|^2},0,\frac{\tau_\ell(\omega_0)-x}{m_2m_3}+\frac{\tau_r(\omega_0)-x}{m_3m_6},0,\cr
    &0,0,\frac{\tau_\ell(\omega_0)-x}{m_4m_7}+\frac{\tau_r(\omega_0)-x}{m_7m_8},0,\frac{\tau_r(\omega_0)-x}{|m_8|^2}\big)^\top.
\end{align}
Therefore, $C_1$ to the second order of perturbation is given by 
\begin{equation*}
    C_1=C^{(0)}+gC^{(1)}+g^2C^{(2)},
\end{equation*}
where $C^{(0)}$, $C^{(1)}$ and $C^{(2)}$ are actively the matricise form of $\ket{C^{(0)}}$, $\ket{C^{(1)}}$ and $\ket{C^{(2)}}$ in Eqs.~(\ref{eq:ketc_0}), (\ref{eq:ketc_1}) and (\ref{eq:ketc_2}), respectively.

\section{Fidelity}
\label{app:fiddelity}
The fidelity between two Gaussian states $\rho_1$ and $\rho_2$ in terms of their covariance matrices is given by \cite{banchi_quantum_2015}
\begin{align}
\label{eq:FidelityDefinition}
	&\mathcal{F}(\rho_1,\rho_2)\cr
    &=\mathcal{F}_0(C_1,C_2)exp(-\frac{1}{4}\hat{\delta d}^\dag(C_1+C_2)^{-1}\hat{\delta d}),
\end{align}
where $\hat{\delta d}=\hat{d}_2-\hat{d}_1$. Here, $\hat{d}_i$s and $C_i$s are respectively displacement vectors and the covariance matrices of $\rho_i$ for $i=1,2$, and $\mathcal{F}_0(C_1,C_2)$ is
\begin{equation}
  \mathcal{F}_0(C_1,C_2)=\frac{F_{\rm{tot}}}{\sqrt[4]{\operatorname{det}(C_1+C_2)}}.
\end{equation}
Here $F_{tot}$ is
\begin{equation}
	F_{\rm{tot}}^4=\operatorname{det}\left[2\left( \sqrt{\mathbb{I}_{2N}-\frac{1}{4}(C_{\rm aux}\mathbb{Z}_N)^{-2}}+\mathbb{I}_{2N}\right)C_{\rm aux}\right],
\end{equation}
with 
\begin{equation}
	C_{\rm aux}=\mathbb{Z}_N(C_1+C_2)^{-1}(\frac{\mathbb{Z}_N}{4}+C_2\mathbb{Z}_N C_1).
\end{equation}
Matrix $\mathbb{Z}_N$ is defined in Eq.~(\ref{eq:Zn}).

\bibliography{OpenBosonicChainExactLocalGlobal}

\end{document}